\documentclass[final,4p,times,twocolumn,numbers]{elsarticle}
\usepackage{amssymb}
\usepackage{lipsum}
\usepackage[T1]{fontenc}
\usepackage{lmodern}
\usepackage[x11names]{xcolor}

\usepackage{graphicx}
\usepackage{booktabs}
\usepackage{tikz}
\usepackage{pgfplots}
\usepackage{pgfplotstable}
\usepackage[skip=2pt,font=normalsize]{subcaption}
\usepackage{adjustbox}
\usepackage{amsmath}
\usepackage{svg}
\usepackage{multirow}
\usepackage{siunitx}
\usepackage{etoolbox} 
\usepackage[outline]{contour} 
\usepackage{textgreek}
\usepackage{csvsimple}
\usepackage{ifthen,calc}
\usepackage{hhline}
\usepackage{wasysym}
\usepackage{cuted}
\usepackage{svg}
\usepackage[colorinlistoftodos]{todonotes}
\usepackage{tabularx}
\usepackage{placeins}
\usepackage{scalerel,amssymb}
\usepackage{makecell}

\newcolumntype{Y}{>{\centering\arraybackslash}X}

\usepgfplotslibrary{ternary}
\usepgfplotslibrary{statistics}

\usepgfplotslibrary{colorbrewer,colormaps}
\usepackage{booktabs,threeparttable}
\usepackage{array} 
\newcolumntype{C}[1]{>{\centering\arraybackslash}p{#1}} 

\usepackage{tikz}

\usetikzlibrary{trees}
\tikzstyle{bag} = [align=center]
\usepgfplotslibrary{groupplots}
\usetikzlibrary{decorations.pathreplacing}
\usetikzlibrary{shapes.geometric, arrows}

\usetikzlibrary{calc}
\usetikzlibrary{external}
\usepgfplotslibrary{fillbetween}

\tikzstyle{startstop} = [rectangle, rounded corners, minimum width=3cm, minimum height=1cm, text centered, text width = 10cm, draw=black, fill=white]

\tikzstyle{process} = [rectangle, minimum width=3cm, minimum height=1cm, text centered, text width = 6cm, draw=black, fill=white, text width = 10cm]

\tikzstyle{arrow} = [ultra thick,->,>=stealth]
\colorlet{teal}{teal!80!black}
\colorlet{cyan}{cyan!60!black}
\colorlet{orange}{orange!80!black}
\colorlet{red}{red!80!white}
\colorlet{lime}{lime!80!black}

\pgfplotsset{cycle list/RdBu}

\pgfplotsset{
  /pgfplots/error bar legend/.style={
    legend image code/.code={
        \draw[sharp plot,mark=-,mark size=5pt,mark repeat=2,mark phase=1,##1]
        plot coordinates { (0.25cm, -0.5cm) (0.25cm,0cm) (0.25cm, 0.5cm) };%
        \draw[mark repeat=0,mark phase=0]
        plot coordinates {(0.25cm,0cm) (0cm,0cm)};%
        \draw[mark repeat=2,mark phase=2]
        plot coordinates {(-0.5cm,0cm)(.5cm,0cm)   
        (1cm,0cm)         
        };%
        }}}
\pgfplotsset{
  /pgfplots/error bar legend/.style={
    legend image code/.code={
        \draw[sharp plot,mark=-,mark size=5pt, mark repeat=2,mark phase=1,##1]
        plot coordinates { (0.3cm, -0.5cm) (0.3cm,0cm) (0.3cm, 0.5cm) };%
        \draw[mark repeat=2, mark size=2pt, mark phase=2,##1]
        plot coordinates {(-0.5cm,0cm) (0.3cm,0cm) (1cm,0cm)};%
        }}}

\pgfplotsset{
    compat=1.18,
    legend image code/.code={
    \draw[mark repeat=2,mark phase=2]
    plot coordinates {(-0.5cm,0cm)(.3cm,0cm)   
    (1cm,0cm)         
    };%
    }
    }

\usepackage{lineno}
\usetikzlibrary{matrix}
\usepackage[hidelinks]{hyperref}
\biboptions{sort&compress}

\journal{Frontiers in Physics}
\begin{document}

\begin{frontmatter}

\title{Radiochromic Film Dosimetry for VHEE and UHDR: Considerations for the CLEAR Facility and Comparisons with Alanine, RPL and Dosimetry Phantoms}

\author[cern,uio]{Vilde F. Rieker\corref{cor1}}
\author[uox]{Joseph J. Bateman\corref{cor2}}
\author[cern]{Laurence Wroe}
\author[hug]{Misael Caloz}
\author[chuv]{Veljko Grilj}
\author[cern]{Ygor Q. Aguiar}
\author[ptb]{Andreas Schüller}
\author[chuv]{Claude Bailat}
\author[cern]{Wilfrid Farabolini}
\author[cern]{Antonio Gilardi}
\author[uox]{Cameron S. Robertson}
\author[cern,uox]{Pierre Korysko}
\author[cern]{Alexander Malyzhenkov}
\author[cern,uio]{Steinar Stapnes}
\author[hug]{Marie-Catherine Vozenin}
\author[cern,uox]{Manjit Dosanjh}
\author[cern]{Roberto Corsini}

\affiliation[cern]{organization={CERN},
            city={Geneva},
            country={Switzerland}}
\affiliation[uio]{organization={University of Oslo},
            city={Oslo},
            country={Norway}}
\affiliation[uox]{organization={University of Oxford},
            city={Oxford},
            country={United Kingdom}}

\affiliation[chuv]{organization={Institute of Radiation Physics, Lausanne University Hospital (CHUV)},
	city={Lausanne},
	country={Switzerland}}
\affiliation[ptb]{organization={Physikalisch-Technische Bundesanstalt (PTB)},
	city={Braunschweig},
	country={Germany}}

\affiliation[hug]{organization={University Hospital of Geneva (HUG)},
	city={Geneva},
	country={Switzerland}}
\cortext[cor1]{vilde.rieker@cern.ch}
\cortext[cor2]{joe.bateman@ucl.ac.uk}
 
\begin{abstract}
Radiochromic films (RCFs) offer valuable two-dimensional dosimetry capabilities with a conceptually simple operating principle that requires minimal investment, making them particularly suitable for very high energy electron (VHEE) FLASH dosimetry where dosimetry standards are currently lacking. However, achieving high-accuracy measurements with RCFs presents significant practical challenges. Without the standardised protocols that ensure reliable outcomes across applications and facilities, knowledge of RCF behaviour becomes essential. This paper identifies common sources of error in RCF preparation, scanning, and processing while proposing specific mitigation strategies to improve accuracy and efficiency. Using our optimised RCF protocol at the CLEAR facility, we demonstrate relative agreement within 5\% compared to alanine dosimeters when measuring Gaussian VHEE beams, establishing a foundation for reliable dosimetry in these advanced radiotherapy applications.
\end{abstract}

\end{frontmatter}




\section{Introduction}
\label{intro}

FLASH radiotherapy (FLASH-RT) has emerged as a promising cancer treatment modality, whereby the radiation is delivered at ultra-high dose rates (UHDRs), $\langle\dot{D}\rangle\geq 40$~Gy/s, eliciting a biological phenomenon known as the FLASH effect. This effect results in enhanced sparing of healthy tissue while maintaining equivalent tumour control compared to conventional dose rates (CONV) ($\langle\dot{D}\rangle\leq 0.03$~Gy/s)~\cite{Vozenin2022TowardsRadiotherapy}. While preclinical evidence has demonstrated the FLASH effect using various radiation types including X-rays, protons, and low-energy electrons ($E$~$\lesssim$~\qty{20}{MeV})~\cite{Gao2022FirstX-rays, Miles2023FLASHX-Rays, Diffenderfer2020DesignSystem, Favaudon2014UltrahighMice, vozenin22}; very high-energy electron (VHEE) beams of energies $E$~$\gtrsim$~\qty{100}{MeV} offer a particularly promising approach due to their superior penetration depth compared to clinical energy electrons~\cite{DesRosiers2000150-250Therapy, Bohlen2021CharacteristicsTargets} and because they are technologically easier to produce at the required intensities and energies as compared to currently available alternatives such as protons and MV photons~\cite{Montay-Gruel2022FLASHBeams}. Therefore, making this a promising modality for treating deep-seated tumours with FLASH-RT~\cite{Bazalova-Carter2015ComparisonPhantom,Schuler2017VeryPPBS,Bohlen20233D-conformalCancer,Bohlen2024VeryPerformances}.

One of the critical challenges impeding the clinical translation of FLASH-RT, especially for pulsed modalities like VHEE, is the development of accurate and reliable active dosimetry. Several studies have shown that ionisation chambers (IC)\textemdash the standard choice for active dosimetry in conventional radiotherapy\textemdash saturate at the UHDR conditions required for FLASH-RT~\cite{mcmanus, martino20, Poppinga2020VHEEConditions}. This dosimetry challenge has launched an area of research adjacent to the radiobiological studies of FLASH-RT. 

The CERN Linear Electron Accelerator for Research (CLEAR) is one of a very limited number of particle accelerators capable of providing VHEE beams at UHDR. CLEAR is a standalone user-facility based at CERN's Meyrin site and has since 2019 been heavily engaged in research related to FLASH-RT using pulsed VHEE beams delivered at UHDR. It offers a flexible range of electron beam parameters in the VHEE regime, including a wide range of mean and instantaneous dose rates\textemdash which makes it a suitable test-bed for both FLASH-RT research for VHEE and active UHDR dosimetry research. For example, the facility has been used by radiobiologists to study the onset of the FLASH effect in plasmids, zebra fish embryos and drosophila larvae~\cite{hannah_plasmid, Kacem2025ModificationProduction, Hart2024DosimetryElectrons}\textemdash while physicists have studied novel methods for active UHDR dosimetry by use of scintillating fibres, screens and fluorescing solutions~\cite{joe_iop, Hart2024PlasticCLEAR, rieker_nima, strathclyde}. All these experiments rely on passive dosimetry for benchmarking and/or dose assessment. 

At CLEAR the standard choice for passive dosimetry has been radiochromic films (RCF) because they provide information about the transverse beam distribution\textemdash an essential capability when working with small and/or inhomogeneous beams. Considerable effort has been dedicated to optimising procedures for the preparation, scanning and processing of RCFs to ensure the highest possible accuracy in dosimetry for facility users. However, achieving this accuracy often requires trading off some efficiency\textemdash and finding the optimal balance between accuracy and efficiency is highly dependent on the number of samples and the time frame of the experiment. This paper provides an overview and quantification of the various effects that influence RCF accuracy\textemdash as well as mitigation strategies particularly aimed at VHEE/UHDR research facilities that depend on and has a high throughput of RCFs for dosimetry. It also aims to thoroughly document and justify the procedure used for passive dosimetry at CLEAR and to compare with results from radiophotoluminescence dosimeters (RPLD), thermoluminescence dosimeters (TLD) and alanine dosimeters (AD).

\section{RCF Dosimetry}
\label{sec:rcf}
RCFs are a type of passive dosimeter commonly used in radiation oncology to assess spatial dose distributions and verify treatment plans~\cite{rcf_history}. They consist of a self-developing active layer that polymerises and darkens upon radiation exposure. For most RCF types this active layer\textemdash a few tens of microns thick\textemdash is sandwiched between two polyester substrates, each around 100 microns thick. These substrates exhibit near tissue and water equivalence, which limits the discrepancy in dose-to-water calibration~\cite{film_tissue}. An RCF dosimetry system (RFDS) consists of a calibration curve that relates the level of RCF darkening to a known dose for a given production lot, a digitiser (such as a commercial flatbed photo scanner) for digitising the RCFs and an RCF processing protocol. However, there are a number of uncertainties in an RFDS that can lead to significant errors in dose evaluation if consistency is not ensured between the calibration and application procedures. The works by Devic and Bouchard \textit{et al.} thoroughly demonstrate the various factors that influence accuracy in RCF dosimetry~\cite{rcf_history, bouchard}. Most of these uncertainties originate from the handling and processing of the RCFs and these can, to varying degrees, be minimised at the cost of a more rigorous and time consuming dosimetry protocol. 

A protocol that is both feasible and sufficiently accurate must be independently established at different accelerator facilities because they typically have different use cases and time constraints. The optimisation of the protocol requires knowledge of the uncertainties related to each part of the process. At CLEAR, we have adapted a protocol that supports high throughput\textemdash with up to 100 separate pieces irradiated in a single day. In the following sections, we will outline the protocol currently in use and elaborate on the lessons, considerations, and trade-offs encountered along the way.

\subsection{Radiochromic Film Models}
\label{sec:rcf_types}
At the CLEAR facility, RCF dosimetry for VHEE UHDR studies involves the use of a variety of Gafchromic\texttrademark~RCF models which are manufactured by Ashland\texttrademark. EBT3/EBT4 RCFs are used for doses up to 10 Gy, EBT-XD for the range 10\textendash \SI{40}{\gray}, MD-V3 for 40\textendash \SI{100}{\gray}, and HD-V2 to cover the 100\textendash \SI{1000}{\gray} range. At CLEAR, all of these RCF models are used for measurements in water, except for HD-V2 RCFs, which are exclusively used for in-air beam profile and dose measurements above \SI{100}{\gray}. The manufacturer states that all these RCF models exhibit a near energy-independent response with <5\SI{5}{\percent} difference in net optical density when exposed at \SI{100}{keV}, \SI{1}{MeV} and \SI{18}{MeV}, and it has been shown experimentally that there is a good agreement between simulations and RCF measurements up to 200 MeV~\cite{Bohlen2021CharacteristicsTargets, Subiel2014DosimetrySimulations, Lagzda2020InfluenceModalities}. Moreover, the manufacturer states that the RCFs are independent of dose-rate with <\SI{5}{\percent} difference in net optical density for \SI{10}{Gy} accumulated at \SI{3.4}{Gy/min} and \SI{0.034}{Gy/min}~\cite{ebt3, ebt4, ebtxd, mdv3, hdv2}. Studies have found that the dose-rate independence can be extended up to instantaneous dose-rates in the order of $10^9~\mathrm{Gy/s}$~\cite{Bazalova-Carter2015ComparisonPhantom, jaccard_RCF_DR, karsch}. 

The atomic composition of the active layer responsible for darkening is similar between the different RCFs. The key differences between the RCF types is related to the crystalline structure of this layer which results in different radiosensitivity~\cite{rcf_composition}. An overview of the RCF models used at CLEAR is provided by Table~\ref{tab:filmtypes}.

\begin{table*}[htb!]
\caption{Properties of the different models of Gafchromic™ RCFs used at CLEAR.}
    \centering
    \begin{tabular}{ c c c c c }
        \toprule
        \makecell{RCF \\model} & \makecell{Scanned\\sample} & \makecell{Cross\\section} & \makecell{Optimal\\(dynamic)\\ range / Gy} & \makecell{CLEAR\\usage} \\ 
        \midrule
        \makecell{EBT3\\ \cite{ebt3}} & \adjustbox{valign=m}{\includegraphics[width=0.15\columnwidth]{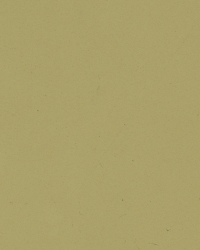}} & 
        \adjustbox{valign=m}{\begin{tikzpicture}[font=\footnotesize]
    \def\layerwidth{4.3cm}
    \def\scale{0.85}
    \node[rectangle, text = black, fill = teal!50, minimum width = \layerwidth, minimum height = \scale*1.25cm, anchor=north west] (top) at (0cm,0cm) {125 \textmu m matte polyester};
    
    \node[rectangle,text = black, fill = red!75, minimum width = \layerwidth, minimum height = \scale*0.28cm, anchor=north west] (mid) at (top.south west)  {28 \textmu m active layer};
  
    \node[rectangle,text = black, fill = teal!50, minimum width = \layerwidth, minimum height = \scale*1.25cm, anchor=north west] (bottom) at (mid.south west) {125 \textmu m matte polyester};
\end{tikzpicture}} & \makecell{0.2\textendash 10\\(0.1\textendash 20)} & \makecell{1\textendash 10 Gy\\in water}\\ \hline
        \makecell{EBT4\\ \cite{ebt4}} & \adjustbox{valign=m}{\includegraphics[width=0.15\columnwidth]{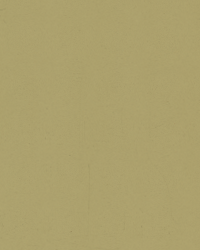}} & 
        \adjustbox{valign=m}{\begin{tikzpicture}[font=\footnotesize]
\def\layerwidth{4.3cm}
\def\scale{0.85}
    \node[rectangle, text = black, fill = teal!50, minimum width = \layerwidth, minimum height = \scale*1.25cm, anchor=north west] (top) at (0cm,0cm) {125 \textmu m matte polyester};
    
    \node[rectangle,text = black, fill = red!75, minimum width = \layerwidth ,minimum height = \scale*0.28cm, anchor=north west] (mid) at (top.south west)  {28 \textmu m active layer};
  
    \node[rectangle,text = black, fill = teal!50, minimum width = \layerwidth, minimum height = \scale*1.25cm, anchor=north west] (bottom) at (mid.south west) {125 \textmu m matte polyester};
\end{tikzpicture}} & 0.2\textendash 10 & \makecell{1\textendash 10 Gy\\in water}\\ \hline
        \makecell{EBT-XD\\ \cite{ebtxd}} & \adjustbox{valign=m}{\includegraphics[width=0.15\columnwidth]{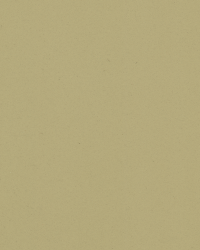}} & 
        \adjustbox{valign=m}{\begin{tikzpicture}[font=\footnotesize]
\def\layerwidth{4.3cm}
\def\scale{0.85}
    \node[rectangle, text = black, fill = teal!50, minimum width = \layerwidth, minimum height = \scale*1.25cm, anchor=north west] (top) at (0cm,0cm) {125 \textmu m matte polyester};
    
    \node[rectangle,text = black, fill = red!70, minimum width = \layerwidth, minimum height = \scale*0.28cm, anchor=north west] (mid) at (top.south west)  {25 \textmu m active layer};
  
    \node[rectangle,text = black, fill = teal!50, minimum width = \layerwidth, minimum height = \scale*1.25cm, anchor=north west] (bottom) at (mid.south west) {125 \textmu m matte polyester};
\end{tikzpicture}} & \makecell{0.4\textendash 40\\(0.1\textendash 60)} & \makecell{10\textendash 40 Gy\\in water}\\ \hline
        \makecell{MD-V3\\ \cite{mdv3}} & \adjustbox{valign=m}{\includegraphics[width=0.15\columnwidth]{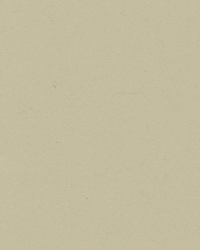}} & 
        \adjustbox{valign=m}{\begin{tikzpicture}[font=\footnotesize]
\def\layerwidth{4.3cm}
\def\scale{0.85}
        \node[rectangle, text = black, fill = teal!50, minimum width = \layerwidth, minimum height = \scale*1.25cm, anchor=north west] (top) at (0cm,0cm) {125 \textmu m matte polyester};
        \node[rectangle,text = black, fill = red!40, minimum width = \layerwidth, minimum height = \scale*0.10cm, anchor=north west] (mid) at (top.south west)  {10 \textmu m active layer};
        \node[rectangle,text = black, fill = teal!50, minimum width = \layerwidth, minimum height = \scale*1.25cm, anchor=north west] (bottom) at (mid.south west) {125 \textmu m matte polyester};
    \end{tikzpicture}} & 1\textendash 100 & \makecell{40\textendash 100 Gy\\in water}\\ \hline
        \makecell{HD-V2\\ \cite{hdv2}} & \adjustbox{valign=m}{\includegraphics[width=0.15\columnwidth]{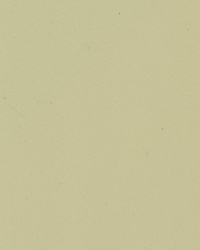}} & 
        \adjustbox{valign=m}{ \begin{tikzpicture}[font=\footnotesize]
        \def\layerwidth{4.3cm}
        \def\scale{0.85}

        \node[rectangle, text = black, fill = none, minimum width = \layerwidth, minimum height = \scale*1.55cm, anchor=north west] (top) at (0cm,0cm) {};
        \node[rectangle,text = black, fill = red!50, minimum width = \layerwidth, minimum height = \scale*0.12cm, anchor = north west] (mid) at (top.south west) {12 \textmu m active layer};
        \node[rectangle,text = black, fill = teal!40, minimum width = \layerwidth, minimum height = \scale*0.97cm, anchor=north west] (bottom) at (mid.south west) {97 \textmu m clear polyester};
    \end{tikzpicture}} & 10\textendash 1000 & \makecell{100\textendash 1000 Gy\\in air}\\ 
        \bottomrule
    \end{tabular}
    
    \label{tab:filmtypes}
\end{table*}

\subsection{RCF Handling and Preparation}
\label{subsec:rcf_prep}
Due to both the physical nature of the RCFs and the method of data retrieval, caution is required in storage and handling to limit possible uncertainties. In particular, it is essential to limit the splitting of the substrates surrounding the active layer, light exposure, the presence of dirt and the formation of scratches to perform reliable dosimetry with RCFs. This section will detail these effects and outline strategies for mitigating them within an RCF protocol.

\subsubsection{RCF Handling}
\label{subsec:handling}
The EBT3, EBT4, EBT-XD, and MD-V3 RCF models gain robustness by sandwiching the active layer between two polyester sheets. This means they can be irradiated in a water phantom for short periods, and be handled with bare hands. The latter, however, can impart fingerprints, grease and dirt that can distort the scanned image and so it is good practice to either grab them by the edges, use gloves, or wipe them clean with an alcohol swab before scanning. On the other hand, the HD-V2 RCFs have their humidity sensitive active layer exposed and therefore cannot be immersed in water and must in general be handled with more care. 

The Gafchromic\texttrademark~RCF models are relatively insensitive to indoor artificial light and can be left exposed for short periods without noticeable effects. However, they are very sensitive to UV light\textemdash even to artificial light containing UV components, which can have a darkening effect on the RCF with prolonged exposure~\cite{light_effect, film_protocol}. It is therefore crucial that the RCFs are kept in an opaque envelope or container when not in use to avoid unnecessary uncertainties. 

Another factor to consider is the temperature stability of the RCFs. According to the manufacturer, EBT3 RCFs are stable up to $60^\circ\mathrm{C}$, but they recommend storing them below $25^\circ\mathrm{C}$~\cite{ebt3}. A study by Trivedi \textit{et al.} showed that storing RCFs in varying temperatures can lead to self-development, effectively shifting their sensitivity~\cite{ambien_temp}. Therefore, it is good practice to store RCFs not only below $25^\circ\mathrm{C}$, but ideally also in a temperature-controlled environment.

\subsubsection{Cutting and Labelling}
\label{subsec:cutandlabel}
The Gafchromic~\texttrademark ~RCFs are manufactured in sheets measuring $203~\mathrm{mm}\times 254~\mathrm{mm}$. At the CLEAR facility, a laser machine (Epilog Fusion Maker 12) has been acquired to systematically cut the RCF sheets into $35~\mathrm{mm} \times 40.5~\mathrm{mm}$ pieces\textemdash a suitable size to capture most of the CLEAR electron beam after it has been scattered through $\simeq 15~\mathrm{cm}$ of water. These dimensions are also tailored to fit into 3D-printed sample holders used for irradiations at CLEAR.

\begin{figure*}[htb!]
\centering
\input{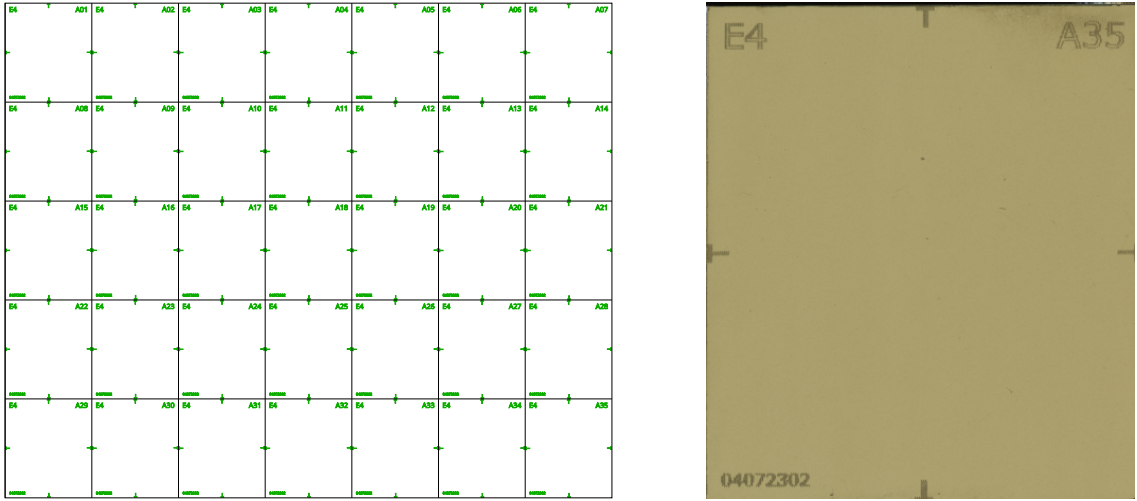}
\caption{\textbf{Left:} SVG-template used for laser cutting and engraving. \textbf{Right:} A cut piece of RCF engraved with the RCF model, LOT number, sample number and alignment markers.}
\label{fig:lasercut}
\end{figure*}

The laser cutting system is also used to engrave and label uniquely the individual RCF pieces. This ensures easy identification and maintenance of RCF orientation during irradiation, scanning, and processing. Examples of a template used for cutting and engraving the RCF sheets and a prepared RCF sample can be seen in Fig.~\ref{fig:lasercut}.

The templates are vector graphics (SVG) files in which the cutting and engraving paths are separated by colour. The Epilog software has a
colour mapping feature which enables the user to assign separate laser attributes to the different colours in the SVG file, which allows for cutting and engraving to be performed in the same process. To prevent damage to the RCFs and achieve satisfactory engraving resolution, it is important to adjust the laser machine's speed, power, and frequency carefully. There is no literature that reports on the specific considerations regarding laser cutting RCFs, but Niroomand-Rad \textit{et al.} state that, if done "carefully", all but a $1~\mathrm{mm}$ margin around the edge can be used for dosimetry~\cite{rcf_composition}. 

The optimal laser settings depend on the specific laser machine used. We have observed, with the Epilog Fusion Maker 12, that cutting at low speeds severely damages the RCF, even at low power. Increasing the speed generally allows for increasing power, and the least visual damage is achieved at high speed and high power. As a remark, it is important to flatten the RFCs in the laser machine by means of e.g. adhesive tape (along the edges outside the engraving area) to keep the laser in focus across the naturally curved sheet.

We explored the possibility of melting the edge of the RCF during the laser cutting process to prevent humidity from entering between the two substrates and diffusing into the active layer. When comparing the combinations of laser settings that did not damage the RCFs, no significant difference in water resistance was observed when cutting at lower or higher power at the same speed. Additionally, no detectable difference in humidity penetration was observed when using 2 cutting cycles at high power compared to using one single cycle. In fact, cutting in multiple cycles is inadvisable, as it may cause the cut pieces to shift due to machine vibrations. 

All cutting settings resulted in diffused layers ranging from 3 to \SI{4}{\milli\metre} into the edges of the RCF pieces after being submerged in water for 10 minutes. It is also important to consider that Gafchromic™ RCFs contain trace amounts of chlorine, as reported by Niiroomand \textit{et al.}~\cite{rcf_composition}. As a safety precaution, it is crucial to cut the RCFs in a well-ventilated area using a laser machine connected to a fume extraction system.

\subsubsection{The RCF preparation protocol at CLEAR}
Based on the considerations in sections~\ref{subsec:handling} and~\ref{subsec:cutandlabel}, the following procedure for RCF preparation has been established at CLEAR:
\begin{enumerate}
    \item RCFs are stored in opaque envelopes at room temperature. 
    \item Gloves are used when handling the RCFs to avoid fingerprints and scratches.
    \item RCF sheets are cut and engraved using a laser machine at high power and high speed, to avoid damage and ensure reproducibility.
  
\end{enumerate}

\subsection{RCF Scanning}
\label{sec:scanning}
RCFs are most commonly processed using 48-bit commercially available flatbed photo scanners in transmission mode. These scanners measure the red, green and blue (RGB) colour components of light transmitted by the RCF at a colour depth of 16 bit per channel. This process yields the fraction of incident light transmitted through the sample $I_t/I_0$ which is mapped to a range of $2^{16}$ pixel values (PV) per color channel from 0 (opaque) to 65535 (transparent). There are several scanning-related factors that can affect the accuracy of RCF dosimetry and this section outlines the effects that exist and the mitigation techniques considered at CLEAR. 

\subsubsection{Timing the Scans}
\label{subsec:timing}
\begin{figure*}[htb!]
    \centering
    \pgfplotsset{compat=1.9}

\begin{tikzpicture}
    \def\Hseparation{2cm}
    \def\Vseparation{1.2cm}
    \begin{axis}[name=dTplot, 
        width=.47\linewidth,
        height=.45\linewidth,
        xlabel=$\Delta t$ / h,
        ylabel=$\Delta PV_\mathrm{red}$ / \%,
        grid=both,
        xmin=0,xmax=300,
        legend pos=north east,
        ]

        \addplot+ [only marks, color=Blue3, line width=.7, mark=+, mark size=3]  table [x = dTh, y = dPV, col sep=comma] {plots/raw_data/dT_HUG/A015_2Gy.csv}; 
        \label{plot:dT_2Gy}

        \addplot+ [only marks, color=Green3, line width=.7, mark=+, mark size=3,]  table [x = dTh, y = dPV, col sep=comma] {plots/raw_data/dT_HUG/A019_4Gy.csv}; 
        \label{plot:dT_4Gy}

        \addplot+ [only marks, color=Yellow2, line width=.7, mark=+, mark size=3,]  table [x = dTh, y =dPV, col sep=comma] {plots/raw_data/dT_HUG/A021_6Gy.csv}; 
        \label{plot:dT_6Gy}

        \addplot+ [only marks, color=DarkOrange1, line width=.7, mark=+, mark size=3,]  table [x = dTh, y =dPV, col sep=comma] {plots/raw_data/dT_HUG/A025_13Gy.csv}; 
        \label{plot:dT_13Gy}
        
        \addplot+ [only marks, color=Red2, line width=.7, mark=+, mark size=3pt,]  table [x = dTh, y =dPV, col sep=comma] {plots/raw_data/dT_HUG/A027_20Gy.csv}; 
        \label{plot:dT_20Gy}
    \end{axis}

    \begin{axis}[
    name=dose_drift, 
    at={($(dTplot.south east) + (1.2*\Hseparation,0)$)},
    anchor=south west,
    height=0.45\linewidth, 
    width=0.47\linewidth,
    xlabel={$\Delta t_\mathrm{appl.}$ / h},
    ylabel=$\Delta D_{\Delta t-24\mathrm{h}}/D_{24\mathrm{h}}$ / \%,
    grid=both,
    xmin=0, xmax=300,
    ymin=-12.5, ymax=14,
    legend style={font=\footnotesize, 
        legend pos =south east},
    ]
    \addplot+ [only marks, color=Blue3, line width=.7, mark=+, mark size=3pt]  table [
    x = dTh, 
    y=dD24, 
    col sep=comma] {plots/raw_data/dT_HUG/A015_2Gy.csv}; 
        \label{plot:dD_2Gy}
        
        \addplot+ [only marks, color=Green3, line width=.7, mark=+, mark size=3pt]  table [
        x = dTh, 
        y =dD24, 
        col sep=comma] {plots/raw_data/dT_HUG/A019_4Gy.csv}; 
        \label{plot:dD_4Gy}
        
        \addplot+ [only marks, color=Yellow2, line width=.7, mark=+, mark size=3pt]  table [
        x = dTh, 
        y =dD24, 
        col sep=comma] {plots/raw_data/dT_HUG/A021_6Gy.csv}; 
        \label{plot:dD_6Gy}
        
        \addplot+ [only marks, color=DarkOrange1, line width=.7, mark=+, mark size=3pt] table [
        x=dTh, 
        y=dD24, 
        col sep=comma] {plots/raw_data/dT_HUG/A025_13Gy.csv};
        \label{plot:dD_13Gy}
        
        \addplot+ [only marks, color=Red2, line width=.7, mark=+, mark size=3pt,]  table [x = dTh, y =dD24, col sep=comma] {plots/raw_data/dT_HUG/A027_20Gy.csv}; 
        \label{plot:dD_20Gy}

        \draw [-, black, thick] (axis cs: 24,-12.5) -- (axis cs: 24,14);

\end{axis}

\matrix [
            draw,
            fill=white,
            matrix of nodes,
            nodes={ minimum size=10pt, minimum width=40pt, font=\footnotesize},
            minimum width=0.38\linewidth,
            anchor=north,
            at={($(dTplot.south east) + (0.6*\Hseparation,-1.5)$)},
        ]{
             \ref{plot:dD_2Gy} 2 Gy & \ref{plot:dD_4Gy} 4 Gy & \ref{plot:dD_6Gy} 6 Gy & \ref{plot:dD_13Gy} 13 Gy & \ref{plot:dD_20Gy} 20 Gy\\
        };
\end{tikzpicture}
    \caption{\textbf{Left:} The relative change in PV for the red channel of an EBT3 RCF as a function of time after exposure $\Delta t$. \textbf{Right:} The dose deviation of the RCFs scanned at different times after exposure, relative to an RCF scanned at $\Delta t_\mathrm{appl.}\simeq \Delta t_\mathrm{calib.}=\SI{24}{\hour}$.}
\label{fig:dT}
\end{figure*}
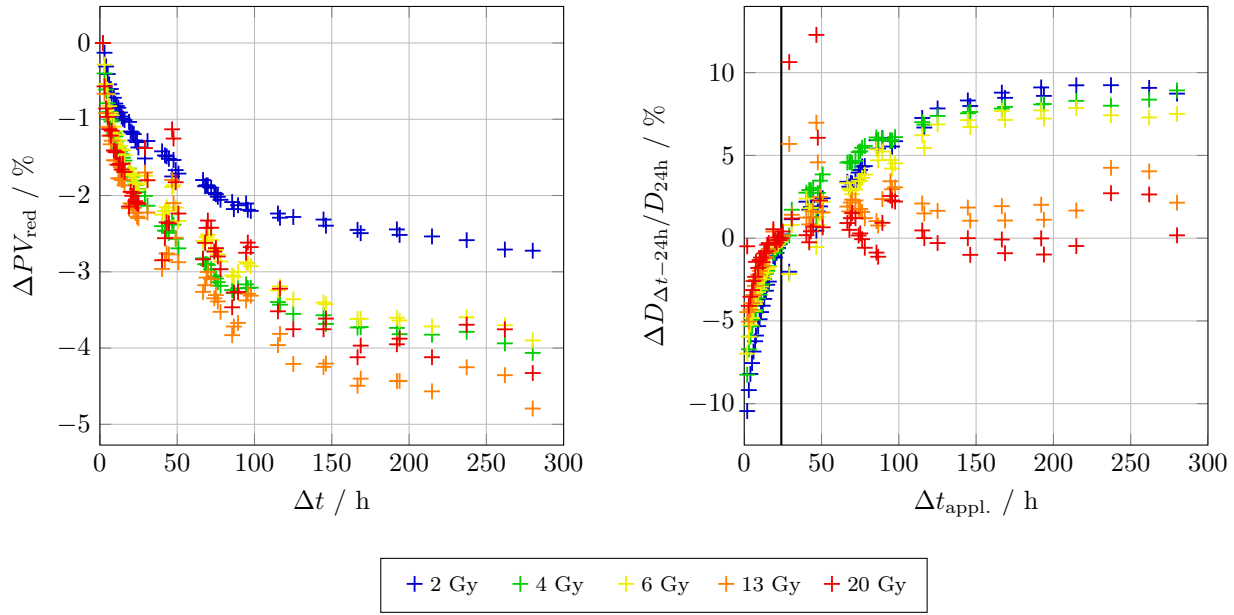

RCFs begin to self-develop immediately upon radiation exposure and the polymerization process never fully ends, but just continues at ever-slower rates. Thus, it is important to maintain consistency in the timing of RCF scanning after exposure. The post-exposure self-development of a set of EBT3 RCFs is shown in Fig.~\ref{fig:dT}.

The results indicate that scanning application RCFs at a $\Delta t_\mathrm{appl.}$ that is significantly different from the $\Delta t_\mathrm{calib.}$ used for the calibration RCFs can result in dose evaluation errors of up to \qty{10}{\percent}. Additionally, the stabilisation appears to occur more rapidly for RCFs exposed to doses at the high end of the RCF's dynamic range. Because the curves are steeper at lower $\Delta t$, the time window to scan without introducing significant error is narrower when scanning early. In other words: the shorter the $\Delta t_\mathrm{calib.}$, the smaller the time-window available to scan the application RCFs without introducing significant dose-errors. Literature thus often suggests to wait for \SI{24}{\hour} post-irradiation before scanning because the polymerization process is more stable at this point. Beyond \qty{24}{\hour} and up to \qty{14}{\day} post exposure, the change is about \qty{2.5}{\percent}~\cite{rcf_composition}.

In summary, the crucial point is to ensure that time of scanning post-exposure of the application RCF ($\Delta t_\mathrm{appl.}$) relative to that of the calibration RCF ($\Delta t_\mathrm{calib.}$) is minimised:
\begin{equation}
    \left|1-\frac{\Delta t_\mathrm{appl.}}{\Delta t_\mathrm{calib.}}\right|\ll 1\ .
    \label{eq:dT_criterion}
\end{equation}

There is a technical possibility of mitigating this timing uncertainty by applying the "one-scan protocol", which also compensates for inter-scan variabilities~\cite{one-scan_micke}. However, this approach relies on a recalibration by irradiating a reference RCF to a known dose similar to the maximum expected value on the application RCF. This is not feasible at CLEAR because we depend on external facilities for RCF calibration. 
 
In practice, particularly at CLEAR where we typically scan large batches of RCFs one-by-one, maintaining a consistent scanning window is challenging. Based on the aforementioned considerations, it is considered best practice to strictly adhere to $\Delta t_\mathrm{appl.}=\Delta t_\mathrm{calib.}\simeq\SI{24}{\hour}$, where the time-window for scanning without significant error is broader. This protocol offers a good compromise between post-exposure stabilisation and acceptable delay for most doses used at CLEAR.
 
\subsubsection{Sample and Scanner Preparation}
\begin{figure*}
    \centering
    \begin{tikzpicture}
\begin{axis}[at={(0,0)},
    name=film,
    xlabel= X / mm,
    xtick={0, 59, 118,176, 235},
    xticklabels={0, 5, 10,15, 20},
    ytick={0, 59, 118,176, 235},
    yticklabels={0, 5, 10,15, 20},
    ylabel=Y / mm,
    xmin=0,
    ymin=0, 
    xmax=235, 
    ymax=235,
    height=0.45\linewidth,
    width=0.45\linewidth,
    axis on top=true,
]
\addplot graphics [
            xmin = 0,
            xmax = 300,
            ymin = 0,
            ymax = 300
        ]
            {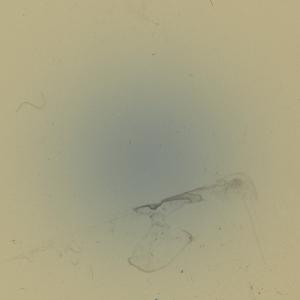};
\draw [-, red, thick] (150,0) -- ++(0,234);
\end{axis}
\begin{axis}[at={(320,0)},
    name=pvplot,
    grid=both, 
    xlabel= Y / mm ,
    ylabel=Dose / Gy,
    xmin=0,
    xmax=235,
    xtick={0, 59, 118,176, 235},
    xticklabels={0, 5, 10,15, 20},
    height=0.45\linewidth, 
    width=0.45\linewidth,
]
    \addplot[color=red]
    table [x=idx, y=Dose_y, col sep=comma] {plots/raw_data/profiles_M004.csv}; 

    \end{axis}

\end{tikzpicture}
    \caption{\textbf{Left:} An example of an RCF with water residues. \textbf{Right:} The corresponding vertical dose profile.}
\label{fig:water_residue}
\end{figure*}

It is strongly advised to warm up the scanner's electronics before starting to scan the RCFs to ensure reproducible results. If the scanner has not been used in the last hour, turning it on at least 30 minutes prior to starting the scans is sufficient. The RCF manufacturer and most studies then suggest to perform at least 5 "preview" scans to warm up the light source and stabilise the response to limit inter-scan variabilities~\cite{ rcf_composition,film_protocol,ipac23_passive}. This is of particular importance at CLEAR where the RCFs are scanned one-by-one. Before scanning the RCFs, it is essential to ensure that both the scanner plate and RCF piece are free from grease, dust and dirt. Both the scanner plate and RCF can be wiped clean with alcohol, and gloves or a vacuum suction pen should be used to position the RCFs on the scanner to keep the sample clean. At CLEAR, where the RCFs are often irradiated in a water phantom, deposits are often observed on the RCF and must be removed before scanning. As illustrated in Fig.~\ref{fig:water_residue}, such artefacts can affect the pixel value and eventually the dose calculation significantly. 

\subsubsection{Consistent Positioning of the RCFs}
The manufacturer recommends the Epson Expression 11000XL Photo scanner because of its large scanning area ($310\times\SI{437}{\milli\metre\squared}$). Large scanning areas are desirable to reduce the impact of lateral response artifacts (LRA) which refer to the systematic dependency of measured PVs on the lateral position of the RCF relative to the centre of the scanner. The LRA results in RCFs scanned closer to the edges of the scanner appearing darker, and the deviation also increases with decreasing PV (i.e. for darker RCFs)~\cite{ rcf_composition, scanner_lra}. 

At CLEAR we have previously found that dose evaluation errors of up to \qty{10}{\percent} can arise from a lateral offset of the application RCF relative to the calibration RCF during scanning~\cite{ipac23_passive}. The LRA is mainly attributed to the fact that the light transmitted through the RCF is polarized by the polymers in the active layer. This light passes via a set of mirrors and a lens before reaching the image sensor, but the angle of incidence increases with the distance from the centre of the scanner~\cite{ebt3}. With a scanner which is large relative to the RFC sample, this effect is limited and can be mitigated through the use of multi-channel processing, discussed in section~\ref{sec:rcf_processing}. 

\begin{figure*}[htb!]
    \centering
    \input{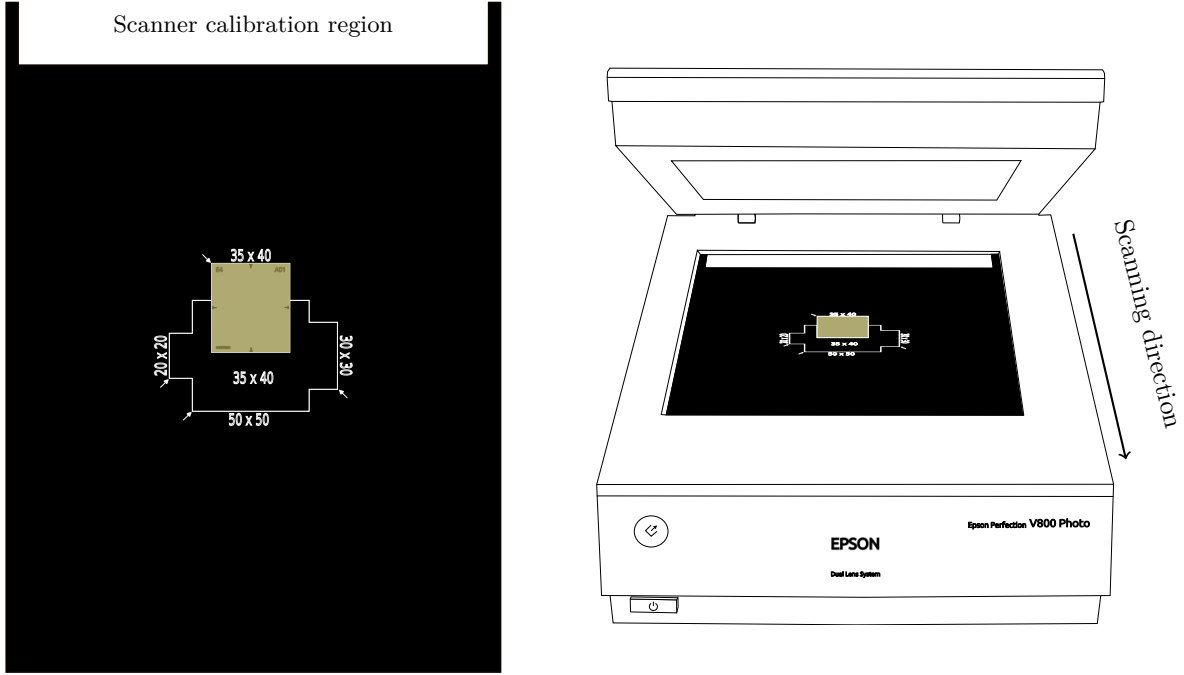}
    \caption{\textbf{Left:} The scanner mask used for reproducible positioning of the RCFs. \textbf{Right:} The Epson Perfection V800 Photo scanner with positioning of the mask in the scanner relative to the scanning direction.}
\label{fig:mask}
\end{figure*}

To eliminate the effect, an LRA correction matrix could also be considered, in particular for higher precision measurements and/or for larger or multiple RCFs~\cite{lewis_lra}. At the CLEAR facility an Epson Perfection V800 Photo scanner, which has a smaller scanning area ($216\times \SI{297}{\milli\metre\squared}$) than the recommended model, is used. However, this is considered acceptable because of the relatively small field sizes\textemdash and thus smaller RCF pieces of $40 \times \SI{35}{\milli\metre\squared}$ used at CLEAR\textemdash which are scanned one-by-one and each centred on the scanner to limit the LRA. Consistent positioning of the RCFs on the scanner is ensured by the use of a custom made scanner mask with different inserts for different RCF sizes as shown in Fig.~\ref{fig:mask}.

It is essential that this mask is positioned consistently on the scanner surface and that the area closest to the scanning direction's starting point is left blank to ensure proper scanner response calibration. 

In addition to ensuring consistent RCF positioning, the scanner mask helps in keeping a consistent orientation of the RCFs throughout the scanning process. It is a well-known fact that the scans are orientation dependent due to the orientation of the polymers in the active layer~\cite{film_protocol,zeidan_orientation, energy_orientation, rcf_scan_considerations}. We have previously reported that increasing relative orientation differences between calibration and application RCFs can yield dose evaluation errors that reach \SI{27}{\percent} at \ang{180} for an EBT3 RCF exposed to \qty{5}{\gray}~\cite{ipac22}. It is therefore crucial to be consistent with the orientation of the application RCFs on the scanner with respect to the orientation for which the calibration RCFs were scanned.   

As a remark, the fact that Gafchromic\texttrademark~RCF sheets are naturally curved may also affect the scanner response. For a full sheet on a flat surface, the maximum distance between the surface and the RCF is in the order of 1\textendash\SI{2}{\milli\metre}, depending on the model and production lot. Particularly for large samples, it is therefore recommended to flatten the RCF on the scanner by means of a clear 3\textendash 4 mm thick glass plate to mitigate the potential error in pixel value reading of \SI{1.2}{\percent} per millimetre offset from the light source~\cite{rcf_composition, film_protocol}. If using a flattening glass plate, it is important to use it consistently for both calibration and application RCFs. At CLEAR, however, we typically cut RCF sheets into small enough pieces (less than \SI{20}{\percent} of the height of the full sheet) to largely mitigate this effect, deeming the consistent use of a glass plate an unnecessary complication.

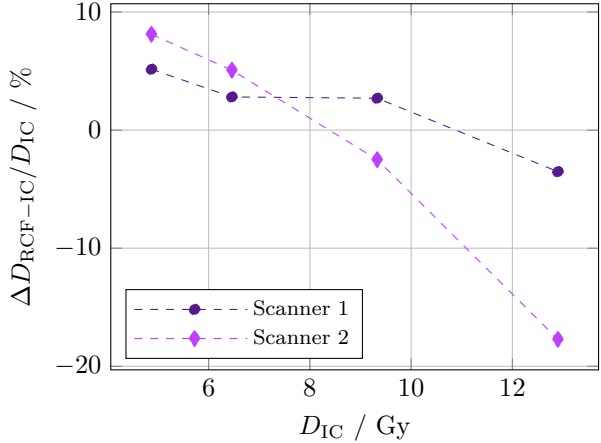
\begin{figure}[htb!]
    \centering
    \begin{tikzpicture}

\begin{axis}[
    name=drift,
    anchor=south west,
    height=0.8\linewidth, 
    width=\linewidth,
    xlabel={$D_\mathrm{IC}$~/~Gy},
    ylabel=$\Delta D_\mathrm{RCF-IC}/D_\mathrm{IC}$ / \%,
    grid=both,
    legend style={font=\footnotesize, 
        legend pos =south west},
    ]

    \addplot+[dashed, color=Purple4, mark=*,mark size=2, mark options={Purple4}, 
        ]
            table [x = IC, y expr= (\thisrow{CLEAR dose}-\thisrow{IC})/\thisrow{IC}*100,  col sep=comma] {plots/raw_data/CLEARvsCHUV.csv}; \label{plot:CLEAR_scan}
    \addlegendentry{Scanner 1}

    \addplot+[dashed, color=DarkOrchid1,mark=diamond*, mark size=3, mark options={DarkOrchid1}, 
        ]
            table [x = IC, y expr=(\thisrow{CHUV dose}-\thisrow{IC})/\thisrow{IC}*100,  col sep=comma] {plots/raw_data/CLEARvsCHUV.csv}; \label{plot:CHUV_scan}
   \addlegendentry{Scanner 2}
    
\end{axis}
\end{tikzpicture}
    \caption{The dose errors of EBT3 RCFs exposed to a known dose $D_\mathrm{IC}$ at a calibration facility and scanned using two different scanners. Scanner 1 refers to the scanner which was also used for the RCF calibration, while scanner 2 is a different scanner of the same type.}
\label{fig:scanners}
\end{figure}

Lastly, it is crucial to always use the same scanner for a given RFDS, although this may present challenges for inter-facility collaborations. For example, it is common for users of the CLEAR facility to bring their own RCFs for experiments. However, since CLEAR lacks RCF calibration capability, users typically rely on their own calibration (from external scanners) to compute doses for application RCFs (scanned at CLEAR) to ensure consistency between  $\Delta t_\mathrm{calib.}$ and $\Delta t_\mathrm{appl.}$. As can be seen in Fig.~\ref{fig:scanners}, using different scanners for the application and calibration RCFs increases the dose error. 
Equally, re-scanning the calibration set at CLEAR will impose a shift of $\Delta t_\mathrm{calib.}$. The easiest way to mitigate this is for users to fully rely on the RFDS at CLEAR, or to construct multiple calibration curves (for various $\Delta t_\mathrm{calib.}$) to allow for more flexibility in scanning time ($\Delta t_\mathrm{appl.}$) for the application RCFs.

\subsubsection{Scanning Software Settings}
It is important to use appropriate scanner settings in the Epson Scan software when digitising RCFs. As per common practise and literature recommendations, transmission mode is used for RCF scanning, with all image corrections features turned off, and the image type set to "48-bit color"~\cite{transmission_reflection}. The manufacturer provides detailed instructions for the Epson Scan software, which are the settings used for scanning at CLEAR~\cite{film_protocol}. Scanned images are stored as TIFF files to maintain adequate bit depth and ensure lossless compression.
In regard to scanner resolution, Gafchromic™~RCFs have a spatial resolution of \SI{25}{\micro\metre} or smaller, and the achievable resolution is limited by the scanning system~\cite{ebt3}. For applications with large and uniform radiation fields, a resolution of 72 dpi (\SI{0.36}{\milli\metre}) is sufficient, and higher resolutions will increase the noise in the measurement~\cite{rcf_uncertainty}. However, for smaller and/or non-uniform fields it may be necessary to increase the resolution to better capture the dose distribution~\cite{rcf_composition}. This adjustment comes at the cost of increased scanning time, image size and processing time. Above all, it is crucial to use the exact same scanner settings are used between application RCFs and calibration RCFs. Given that the typical VHEE beam at CLEAR in some cases has a Gaussian distribution with 1\textsigma~sizes of $\sim 1~\mathrm{mm}$, the RCFs are consistently scanned at a resolution of 300 dpi ($0.085~\mathrm{mm}$), in order to ensure that the distribution is captured accurately in all cases. 

\subsubsection{Scanning Repetitions}
If the one-scan protocol mentioned in section~\ref{subsec:timing} is not employed, it is suggested to perform repeat scans of each RCF sample and evaluate the average in order to mitigate inter-scan variabilities~\cite{rcf_uncertainty}. However, this is a time-consuming task when a large number of RCF samples are used, as is typically the case at CLEAR. Lewis and Devic have instead proposed to mitigate this by including a piece of unexposed RCF in every scan to act as a reference for the scanner response~\cite{inter-scan_correction}. This is a mitigation strategy which is simple and straight forward to implement, and can be used simultaneously for background evaluation and eventual subtraction, as discussed in section~\ref{sec:rcf_calibration}. As long as the reference is properly aligned with the application RCF in the scanning direction, there should be no LRA difference to compensate for. Additionally, studies have shown that repeated scans of an unexposed RCF do not cause any permanent darkening, which means that using such a reference repeatedly should not induce artificial offsets~\cite{one-scan_micke, inter-scan_correction}.

\subsubsection{Summary: The CLEAR Scanning Protocol}
Taking into account all the considerations concerning RCF scanning, we have developed a protocol that optimizes accuracy while maintaining the necessary efficiency for use at CLEAR:

\begin{enumerate}
  \item Ensure all calibration and application RCFs are scanned at the same time $\Delta t\geq24~\mathrm{h}$ post exposure.
  \item Switch on the scanner \qty{30}{\minute} before starting the scanning to warm up the scanner electronics.
  \item Clean the scanner surface and RCFs if they are dirty.
  \item Perform 5 preview scans to warm up the scanner lamp and stabilise the scanner response.
  \item Position the RCFs at the centre of the scanner and with consistent orientation.
  \item Ensure the scanner settings are consistent and without colour corrections. 
\end{enumerate}

Proposed future improvements include consistently scanning a reference RCF, and potentially investing in a larger scanner and/or implementing an LRA correction matrix. The latter will allow for scanning more RCFs simultaneously for increased efficiency and reduced inter-scan variability between samples, even if a one-scan re-calibration procedure remains unfeasible.

\subsection{RCF Calibration}
\label{sec:rcf_calibration}
Each production lot of RCFs must be calibrated to establish the relationship between the level of darkening and absorbed dose. Since there are no established VHEE reference beams, external clinical low-energy electron beams are used for RCF calibration. Previous VHEE dosimetry studies have reported agreement between RCF dose measurements and Monte Carlo simulations whereby the RCFs were calibrated to clinical 15\textendash 20 MeV electron beams~\cite{Subiel2014DosimetrySimulations, Lagzda2020InfluenceModalities, Bohlen2021CharacteristicsTargets}, meaning that a low-energy RCF calibration is valid for VHEE applications. 

\begin{figure*}[htb!]
    \input{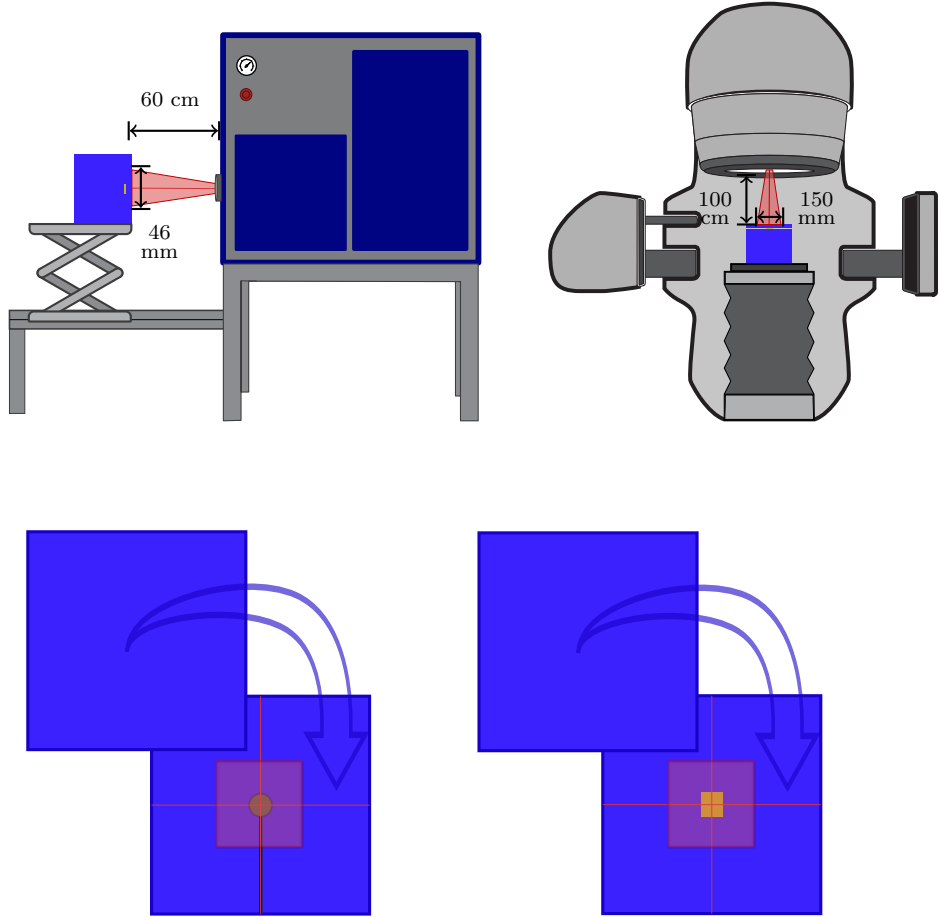}
    \caption{\textbf{Top:} Calibration setup at the eRT6 linac at CHUV (left) and at the Varian TrueBeam medical linac at HUG (right). \textbf{Bottom:} Ionisation chamber (left) and RCF (right) positioned behind $1~\mathrm{cm}$ of solid water.}
    \label{fig:calib_setup}
\end{figure*}

RCFs used in CLEAR are typically calibrated at the Oriatron eRT6 linac at Lausanne University Hospital (CHUV)~\cite{Jaccard2018HighUse}, or in a Varian TrueBeam medical linac at the Geneva University Hospitals (HUG). Both calibration setups provide a 6 MeV electron beam and calibrate the RCFs to a secondary standard ionisation chamber that is placed behind a \SI{1}{\centi\metre} slab of solid water. The ionisation chambers are calibrated as dose to water and traceable to the primary standard at the Swiss Federal Institute of Metrology (METAS). The type of ionisation chamber, the field shape and size, and the temporal beam structure differs between the two setups. For example, the eRT6 has a circular field while the TrueBeam field is square and Table \ref{tab:calib_facilities} presents a comparison of the remaining parameters for the two facilities used. A corresponding illustration of the calibration setups is presented in Fig.~\ref{fig:calib_setup}.

\begin{table*}[htb!]
    \centering
    \caption{The beam parameters used for RCF calibration at eRT6 at CHUV and the Varian TrueBeam medical linac at HUG.}
    \label{tab:calib_facilities}
    \begin{tabularx}{\textwidth}{@{\hskip 0.5cm}X|@{\hskip 0.5cm}Y@{\hskip 0.5cm}Y@{}}\toprule
         \textbf{Parameter} & \textbf{eRT6 (CHUV)} & \textbf{Varian TrueBeam (HUG)} \\  \midrule
         Energy / MeV & 6 & 6 \\
         SSD / cm & 60 & 100 \\
         Beam size / \unit{\milli\metre} & 46 (1\textsigma)& 150 (FWHM)\\
         Dose rate / \unit{\gray\per\second} & 0.14 & 0.16\\
         Pulse width / \unit{\milli\second} & 1 & 4-6\\
         Pulse frequency / \unit{\hertz} & 10 & $\sim 367$\\
         Ionisation chamber & Adv. Markus & Roos\\
         $U_{95,D_\mathrm{calib.}}$ / \unit{\percent} & 3.0 & 2.7\\ 
         Solid water & PTW RW3 & PTW RW3\\
         \bottomrule
    \end{tabularx}   
\end{table*}

RCF calibration requires exposing the separate film pieces to a sufficient number of known doses within the dynamic range of the specific RCF type to establish a good calibration curve. The required number of dose-points depends on the type of fitting function and the desired range of the calibration curve. According to the manufacturer, 6-8 dose-points $D_{\mathrm{calib.},n}$ in geometric progression (i.e. $D_{\mathrm{calib.},n}=D_{\mathrm{calib.},1} r^{n-1}$) is sufficient when using a rational calibration function such as Equation~\ref{eq:calib_func} because this type of function captures the physical nature of the RCF development~\cite{ebt3, ebtxd}. The common ratio $r$ can be determined based on the dose range of interest and, and the desired number of calibration points. Perhaps more importantly, multiple film pieces should be calibrated to the same dose in order to enable meaningful uncertainty estimations.

For the RCF calibrations undertaken for CLEAR, we typically use a couple of additional dose points (8-12 in total) to cover the entire dynamic range of the RCF. For each dose point, two RCF pieces are calibrated simultaneously by stacking them within the solid water phantom. This approach adds statistical robustness without increasing calibration time through separate irradiations and minimises gradients between the two RCFs. If possible, a higher number of samples per dose point can be utilized to improve statistical accuracy and uncertainty estimates. Additionally, including extra dose points between those used for the calibration curve is good practice, as it helps evaluate the curve's estimation accuracy.

\subsubsection{Calibration Response Variables and Fitting Functions}
The response variable to which the dose dependency of the RCF lot is established must be chosen before a calibration curve is constructed. From the TIFF files of the scanned RCFs, one needs to extract the three 2D arrays corresponding to the response of the red, green and blue channels. For each colour, a separate curve relating dose to a response variable must be established. 

There are three response variables that recur in literature; PV, optical density (OD) and net optical density (nOD). PV is generally the choice for simplicity, because the values are obtained directly from the 2D images from each colour channel $x$. The more widespread choice is OD, which relates to PV as
\begin{equation}
\mathit{OD}_x=-\mathrm{ln}\left(\frac{\mathit{PV}_x}{65535}\right),
\label{eq:ODx}
\end{equation}
where 65535 is the RGB colour space of the scanner as described in section~\ref{sec:scanning}. The reason for the conversion to OD is that this quantity exhibits a more linear behaviour as a function of dose compared to PV  and requires fewer non-linear terms to fit the inherently non-linear RCF response to radiation~\cite{rcf_history}. One may extend this further to nOD by subtracting the mean OD of the unexposed RCF $\langle\mathit{OD}_{x,0}\rangle$ from the OD of the exposed RCF $\mathit{OD}_{x}$,

\begin{equation}
\mathit{nOD}_x=\mathit{OD}_{x}-\langle\mathit{OD}_{x,0}\rangle\ .
\label{eq:nODx}
\end{equation}

Since there is not a pixel-to-pixel correspondence between the unexposed and exposed RCFs and the fields are non-uniform, the average OD across a region of interest (ROI) of the unexposed RCF is extracted in order to generate the nOD map for each RCF. This method accounts for the drift of the RCF lot and is therefore recommended~\cite{transmission_reflection}. Ideally, the background should be obtained from the same RCF piece before irradiation. However, due to the number of RCF pieces used at CLEAR and the fact that the films are scanned one-by-one, one RCF piece is typically kept from each sheet to use as background for all irradiated RCFs from that sheet. 

There are numerous variants of fitting functions in literature; however it is key to choose a function which most accurately expresses the RCFs physical response to radiation and fits the calibration points. The darkening of RCFs increases with increased exposure, yet it approaches a near constant value towards the end of the dynamic range (saturation). To this end, there are generally three families of functions that express this behaviour: rational, polynomial and exponential functions. Devic et al. showed that the  choice may depend on the specific RFDS~\cite{devic_digitizers}. More generally speaking, the performance of each function will depend on the number of calibration dose points $D_\mathrm{calib.}$ and dose-range. 

The manufacturer recommends rational functions of the form proposed by Micke et al.~\cite{micke_multi}
\begin{equation}
\mathit{nOD}_x(D) = -\mathrm{ln}\left(\frac{a_x + b_xD}{c_x + D}\right),
\label{eq:calib_func}
\end{equation}
where $a_x$, $b_x$ and $c_x$ are fitting parameters for colour channel $x$. Such functions are best aligned with the inherent behaviour of the RCF, as opposed to polynomial functions, which do not necessarily correspond to the physical properties of the RCFs outside the range of the fitted data points~\cite{hdv2}. 

\begin{figure*}[htb!]
    \centering
     \input{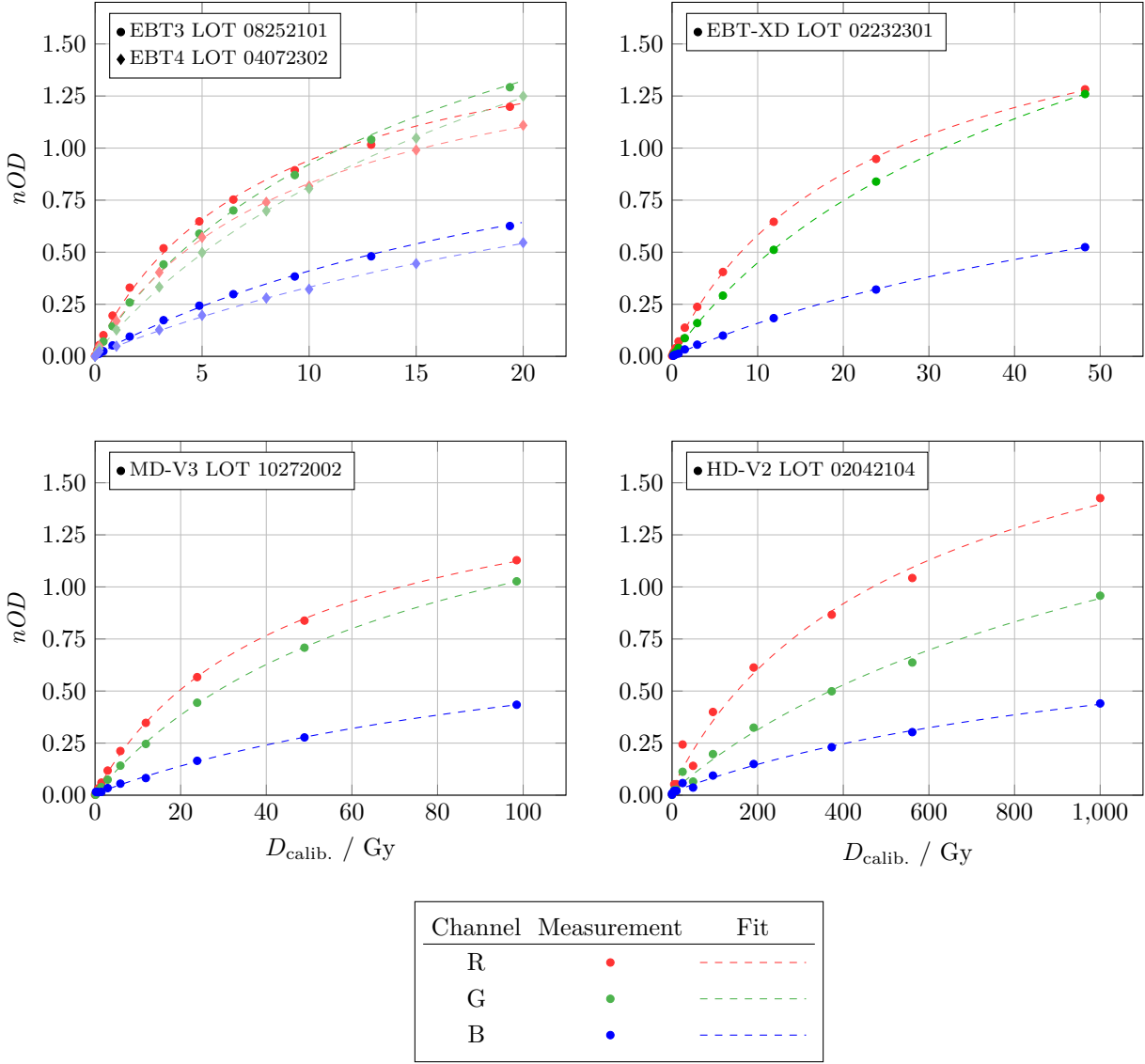}
     \caption{Calibration curves for different RCF models.}
     \label{fig:calib_curves}
\end{figure*}

At CLEAR we have typically used the fitting function in Equation~\ref{eq:calib_func} and Fig.~\ref{fig:calib_curves} shows fitted calibration curves for the different RCF models. Visually, it is evident that the Equation~\ref{eq:calib_func} is a good fit to the dose points of the EBT3, EBT-XD, and MD-V3 RCFs with the fit lying within \qty{0.2}{\percent} of the IC measurement $D_\mathrm{calib.}$, which is less than the uncertainty $U_{95,D_\mathrm{calib.}}$. The exception is HD-V2 for which we speculate that the large uncertainty is due to  environmental exposure or handling\textemdash to which they are more sensitive, as mentioned in section~\ref{sec:rcf_types}. However, because they are less frequently used, we do not have enough statistics to verify this.

\subsubsection{Establishing the Calibration Curves}
\begin{figure*}[htb!]
    \centering
    \input{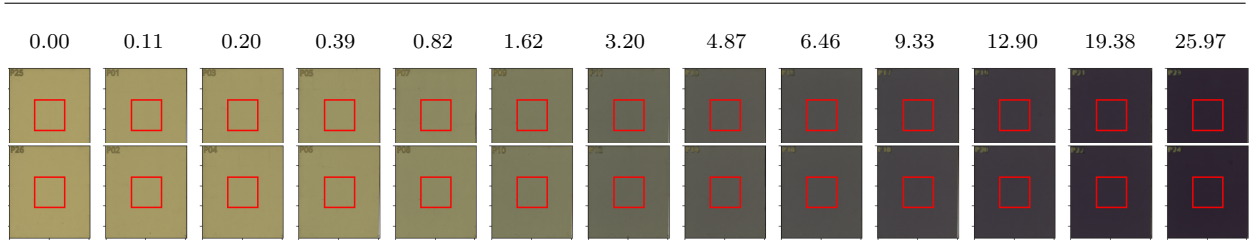}
    \caption{A scanned calibration set of EBT3 RCFs with two samples per dose point and the ROI used for response evaluation indicated.}
    \label{fig:calib_set}
\end{figure*}

A calibration set of EBT3 RCFs over is shown in Fig.~\ref{fig:calib_set}. 
Each RCF has been exposed to a uniform field of known doses provided by the ionisation chamber at the calibration facility. 

To extract the calibration curve, an ROI must be determined that: omits the edges and engravings of the RCF, is fully covered by the uniform field, and is large enough that a histogram of all PVs result in a normal distribution with a measurable standard deviation~\cite{rcf_composition}.

For this case, an ROI of $150\times150$ pixels about the RCF centre was selected. This corresponds to an ROI of $12.75\times12.75~\mathrm{mm^2}$ when scanned at $300$~dpi. For each dose point, the PV distribution is converted to OD using Eq.~\ref{eq:ODx} and then to nOD using Eq.~\ref{eq:nODx}\textemdash which includes a \textit{background subtraction} by the mean OD of the unexposed (\SI{0}{Gy}) RCF. The response of each RCF is the mean nOD within the ROI, and the response of each dose point is the mean response between the two RCF pieces exposed to the same dose.

\subsubsection{Uncertainty estimation}
\label{calib_uncertainty}
The uncertainty in measured dose using RCFs  is a combination of uncertainties originating from the calibration dose $D_\mathrm{calib.}$, inherent RCF imperfections, scanner response, calibration function and the protocol of the RFDS. Estimating the uncertainty for a given calibration curve is outlined by Devic et al. where the error in dose response, $\sigma_\mathit{nOD_x}$, is expressed via error propagation from $\sigma_\mathit{PV_x}$ via $\sigma_\mathit{OD_x}$~\cite{devic_digitizers, devic_precise}.

Converting individual PVs to OD gives an uncertainty
\begin{equation}
    \sigma_{\mathit{OD_x}}=\sqrt{\left(\frac{\partial \mathit{OD_x}}{\partial \mathit{PV_x}}\right)^2\sigma_{\mathit{PV_x}}^2}=\frac{\sigma_{\mathit{PV_x}}}{\mathit{PV_x}^2}\ .
\end{equation}
Propagating this further, the uncertainty of the $\mathit{nOD}$ becomes
\begin{equation}
    \sigma_{\mathit{nOD_x}}=\sqrt{\left(\frac{\partial \mathit{nOD_x}}{\partial \mathit{OD_x}}\right)^2\sigma_{\mathit{OD_x}}^2+\left(\frac{\partial \mathit{nOD_x}}{\partial \langle\mathit{OD}_{x,0}\rangle}\right)^2\sigma_{\langle\mathit{OD}_{x,0}\rangle}^2}\ ,
    \label{eq:sigma_nod}
\end{equation}
where $\langle\mathit{OD}_{x,0}\rangle$ is the mean OD of the unexposed RCF piece with uncertainty
\begin{equation}
    \sigma_{\langle\mathit{OD}_{x,0}\rangle}=\frac{\sigma_{\mathit{OD}_{x,0}}}{\sqrt{n}}\ ,
\end{equation}
where $n$ is the number of pixels used for estimating the background. 

Similarly, each dose point in the calibration curve is related to a mean nOD value with uncertainty
\begin{equation}
    \sigma_{\langle\mathit{nOD}_x\rangle}=\frac{\sigma_{\mathit{nOD}_x}}{\sqrt{n}}\ .
\end{equation}
Orthogonal distance regression (ODR) calculates the perpendicular distance from the data points to the fitted line\textemdash and is used to take into account uncertainties in both calibration dose ($\sigma_{D_\mathrm{calib.}}$) and nOD ($\mathit{\sigma_\mathit{nOD}}$) to construct the calibration curve~\cite{Virtanen2020SciPyPython}. 
\subsubsection{Calibration Performance}
As stated in Section~\ref{sec:rcf_calibration}, calibrating each lot of RCFs\textemdash even if they are of the same model\textemdash is crucial to account for inter-lot variability. Fig.~\ref{fig:EBT3_curves} shows the different calibration curves yielded by different lots of EBT3 RCFs. Applying a calibration curve from a different lot may lead to dose estimation errors in the order of \SI{20}{\percent}.

\begin{figure*}[htb!]
    \centering
     \pgfplotsset{compat=1.9}
\tikzset{
  mynode/.style={
  , align=center
  , execute at begin node=\setlength{\baselineskip}{1em}
  }
}

\begin{tikzpicture}
\def\Hseparation{1.5cm}
\def\Vseparation{1.2cm}

\begin{axis}[
    name=ebt3_lots, 
    anchor=south west,
    height=0.4\linewidth, 
    width=0.5\linewidth,
    ylabel={$\mathit{nOD}$},
    xlabel={$D_\mathrm{calib.}$~/~Gy},
    grid=both,
    xmin=0, xmax=20,
    ymin=0, ymax=1.4,
    yticklabel style={
      /pgf/number format/fixed,
      /pgf/number format/fixed zerofill,
      /pgf/number format/precision=2
    },
    ytick={0,0.25,0.5,0.75,1,1.25},
    legend style={font=\footnotesize, 
        legend pos =north west},
    ]

    \addplot+[only marks,mark=square*,mark size=1.5, mark options={Brown4}, 
        ]
            table [x = dose, y = red,  col sep=comma,restrict x to domain=0:20] {plots/raw_data/calibration/EBT3_08252101_19-10-2022_eRT6_nOD_calibdata.csv}; \label{plot:EBT3_L1_red_raw}

    \addplot [dashed, color=Brown4,  forget plot
      ]  table [x = fit_doses, y = fit_red, col sep=comma,restrict x to domain=0:20] {plots/raw_data/calibration/EBT3_08252101_19-10-2022_eRT6_nOD_calibdata.csv}; \label{plot:EBT3_L1_red_fit}
    
    \addplot+[only marks,mark=square*,mark size=1.5, mark options={Chartreuse4}, forget plot
            ]  
            table [x = dose, y = green, col sep=comma, restrict x to domain=0:20] {plots/raw_data/calibration/EBT3_08252101_19-10-2022_eRT6_nOD_calibdata.csv}; \label{plot:EBT3_L1_green_raw}
    
    \addplot [dashed, color=Chartreuse4, forget plot
      ]  table [x = fit_doses, y = fit_green, col sep=comma,restrict x to domain=0:20] {plots/raw_data/calibration/EBT3_08252101_19-10-2022_eRT6_nOD_calibdata.csv}; \label{plot:EBT3_L1_green_fit}
    
    \addplot+[only marks,mark=square*,mark size=1.5, mark options={Blue3}, 
            forget plot]  
            table [x = dose, y = blue,   col sep=comma,restrict x to domain=0:20] {plots/raw_data/calibration/EBT3_08252101_19-10-2022_eRT6_nOD_calibdata.csv}; \label{plot:EBT3_L1_blue_raw}
    
    \addplot [dashed, color=Blue3, forget plot
      ]  table [x = fit_doses, y = fit_blue, col sep=comma,restrict x to domain=0:20] {plots/raw_data/calibration/EBT3_08252101_19-10-2022_eRT6_nOD_calibdata.csv}; \label{plot:EBT3_L1_blue_fit}

    \addplot+[only marks,mark=diamond*, mark options={Brown3}, 
        ]
            table [x = dose, y = red,  col sep=comma,restrict x to domain=0:21] {plots/raw_data/calibration/EBT3_09302201_07-11-2024_eRT6_nOD_calibdata.csv}; \label{plot:EBT3_L2_red_raw}
    
    \addplot [dashed, color=Brown3,  forget plot
      ]  table [x = fit_doses, y = fit_red, col sep=comma,restrict x to domain=0:21] {plots/raw_data/calibration/EBT3_09302201_07-11-2024_eRT6_nOD_calibdata.csv}; \label{plot:EBT3_L2_red_fit}
    
    \addplot+[only marks,mark=diamond*, mark options={SeaGreen4}, forget plot
            ]  
            table [x = dose, y = green, col sep=comma, restrict x to domain=0:21] {plots/raw_data/calibration/EBT3_09302201_07-11-2024_eRT6_nOD_calibdata.csv}; \label{plot:EBT3_L2_green_raw}
    
    \addplot [dashed, color=SeaGreen4, forget plot
      ]  table [x = fit_doses, y = fit_green, col sep=comma,restrict x to domain=0:21] {plots/raw_data/calibration/EBT3_09302201_07-11-2024_eRT6_nOD_calibdata.csv}; \label{plot:EBT3_L2_green_fit}
    
    \addplot+[only marks,mark=diamond*, mark options={RoyalBlue3}, forget plot
            ]  
            table [x = dose, y = blue,   col sep=comma,restrict x to domain=0:21] {plots/raw_data/calibration/EBT3_09302201_07-11-2024_eRT6_nOD_calibdata.csv}; \label{plot:EBT3_L2_blue_raw}
    
    \addplot [dashed, color=RoyalBlue3, forget plot
      ]  table [x = fit_doses, y = fit_blue, col sep=comma,restrict x to domain=0:21] {plots/raw_data/calibration/EBT3_09302201_07-11-2024_eRT6_nOD_calibdata.csv}; \label{plot:EBT3_L2_blue_fit}

    \addplot+[only marks,mark=*, mark options={IndianRed3}, mark        size=1.5
        ]
            table [x = dose, y = red,  col sep=comma,restrict x to domain=0:21] {plots/raw_data/calibration/EBT3_03022104_09-03-2022_eRT6_nOD_calibdata.csv}; \label{plot:EBT3_L3_red_raw}
    
    \addplot [dashed, color=IndianRed3,  forget plot
      ]  table [x = fit_doses, y = fit_red, col sep=comma,restrict x to domain=0:21] {plots/raw_data/calibration/EBT3_03022104_09-03-2022_eRT6_nOD_calibdata.csv}; \label{plot:EBT3_L3_red_fit}
    
    \addplot+[only marks,mark=*, mark size=1.5, mark options={PaleGreen3}, forget plot
            ]  
            table [x = dose, y = green, col sep=comma, restrict x to domain=0:21] {plots/raw_data/calibration/EBT3_03022104_09-03-2022_eRT6_nOD_calibdata.csv}; \label{plot:EBT3_L3_green_raw}
    
    \addplot [dashed, color=PaleGreen3, forget plot
      ]  table [x = fit_doses, y = fit_green, col sep=comma,restrict x to domain=0:21] {plots/raw_data/calibration/EBT3_03022104_09-03-2022_eRT6_nOD_calibdata.csv}; \label{plot:EBT3_L3_green_fit}
    
    \addplot+[only marks,mark=*, mark size=1.5, mark options={RoyalBlue1}, 
            forget plot]  
            table [x = dose, y = blue,   col sep=comma,restrict x to domain=0:21] {plots/raw_data/calibration/EBT3_03022104_09-03-2022_eRT6_nOD_calibdata.csv}; \label{plot:EBT3_L3_blue_raw}
    
    \addplot [dashed, color=RoyalBlue1, forget plot
      ]  table [x = fit_doses, y = fit_blue, col sep=comma,restrict x to domain=0:21] {plots/raw_data/calibration/EBT3_03022104_09-03-2022_eRT6_nOD_calibdata.csv}; \label{plot:EBT3_L3_blue_fit}

    \addplot+[only marks,mark=triangle*, mark options={red!40!white}
        ]
            table [x = dose, y = red,  col sep=comma,restrict x to domain=0:21] {plots/raw_data/calibration/EBT3_03122004_19-10-2022_eRT6_nOD_calibdata.csv}; \label{plot:EBT3_L4_red_raw}
    
    \addplot [dashed, color=red!40!white,  forget plot
      ]  table [x = fit_doses, y = fit_red, col sep=comma,restrict x to domain=0:21] {plots/raw_data/calibration/EBT3_03122004_19-10-2022_eRT6_nOD_calibdata.csv}; \label{plot:EBT3_L4_red_fit}
    
    \addplot+[only marks,mark=triangle*, mark options={green!50!black!30!white}, forget plot
            ]  
            table [x = dose, y = green, col sep=comma, restrict x to domain=0:21] {plots/raw_data/calibration/EBT3_03122004_19-10-2022_eRT6_nOD_calibdata.csv}; \label{plot:EBT3_L4_green_raw}
    
    \addplot [dashed, color=green!50!black!30!white, forget plot
      ]  table [x = fit_doses, y = fit_green, col sep=comma,restrict x to domain=0:21] {plots/raw_data/calibration/EBT3_03122004_19-10-2022_eRT6_nOD_calibdata.csv}; \label{plot:EBT3_L4_green_fit}
    
    \addplot+[only marks,mark=triangle*, mark options={blue!40!white}, 
            forget plot]  
            table [x = dose, y = blue,   col sep=comma,restrict x to domain=0:21] {plots/raw_data/calibration/EBT3_03122004_19-10-2022_eRT6_nOD_calibdata.csv}; \label{plot:EBT3_L4_blue_raw}
    
    \addplot [dashed, color=blue!40!white, forget plot
      ]  table [x = fit_doses, y = fit_blue, col sep=comma,restrict x to domain=0:21] {plots/raw_data/calibration/EBT3_03122004_19-10-2022_eRT6_nOD_calibdata.csv}; \label{plot:EBT3_L4_blue_fit}

\end{axis}

\begin{axis}[
    name=lot_drift, 
    at={($(ebt3_lots.south east) + (1.2*\Hseparation,0)$)},
    anchor=south west,
    height=0.4\linewidth, 
    width=0.5\linewidth,
    xlabel={$D_\mathrm{calib.}$~/~Gy},
    ylabel=$\Delta D_\mathrm{lot_n-lot_1}/D_\mathrm{lot_1}$ / \%,
    grid=both,
    xmin=2, xmax=21,
    ]
    \addplot+[dashed, color=black,mark=square*, mark size=2, mark options={black}, 
        ]  
        table [x = IC, y expr = (\thisrow{false dose}-\thisrow{real dose})/\thisrow{real dose}*100,   col sep=comma] {plots/raw_data/calibration/lot_08252101.csv}; \label{plot:lot1_dose}
    
    \addplot+[dashed, color=DarkOrchid1,mark=diamond*, mark size=3, mark options={DarkOrchid1}, 
        ]  
        table [x = IC, y expr = (\thisrow{false dose}-\thisrow{real dose})/\thisrow{real dose}*100,   col sep=comma] {plots/raw_data/calibration/lot_09302201.csv}; \label{plot:lot2_dose}

    \addplot+[dashed, color=Purple4,mark=*, mark size=2, mark options={Purple4}, 
        ]  
        table [x = IC, y expr = (\thisrow{false dose}-\thisrow{real dose})/\thisrow{real dose}*100,   col sep=comma] {plots/raw_data/calibration/lot_03022104.csv}; \label{plot:lot3_dose}

    \addplot+[dashed, color=SlateBlue2,mark=triangle*, mark size=3, mark options={SlateBlue2}, 
        ]  
        table [x = IC, y expr = (\thisrow{false dose}-\thisrow{real dose})/\thisrow{real dose}*100,   col sep=comma] {plots/raw_data/calibration/lot_03122004.csv}; \label{plot:lot4_dose}
\end{axis}

\matrix [
            draw,
            fill=white,
            matrix of nodes,
            nodes={ minimum size=15pt, minimum width=40pt},
            anchor=north,
            at={($(ebt3_lots.south east) + (0.6*\Hseparation,-1.7)$)},
        ] {
             LOT no. & 1 & 2 & 3 & 4\\
            \hline \\
            & \ref{plot:EBT3_L1_red_raw} \ref{plot:EBT3_L1_green_raw} \ref{plot:EBT3_L1_blue_raw}& \ref{plot:EBT3_L2_red_raw} \ref{plot:EBT3_L2_green_raw} \ref{plot:EBT3_L2_blue_raw} &              \ref{plot:EBT3_L3_red_raw} \ref{plot:EBT3_L3_green_raw} \ref{plot:EBT3_L3_blue_raw} &              \ref{plot:EBT3_L4_red_raw} \ref{plot:EBT3_L4_green_raw} \ref{plot:EBT3_L4_blue_raw} \\
             & \ref{plot:lot1_dose} & \ref{plot:lot2_dose} & \ref{plot:lot3_dose} & \ref{plot:lot4_dose}\\
        };

\end{tikzpicture}
     \caption{\textbf{Left:} Calibration curves for different production lots of EBT3 RCFs. \textbf{Right:} The relative errors induced by applying the calibration curve of LOT 1 to the other lots.}
     \label{fig:EBT3_curves}
\end{figure*}

Moreover, as discussed in Section~\ref{subsec:timing}, it is important to be consistent with the post-exposure timing between calibration ($\Delta t_\mathrm{calib.}$) and application ($\Delta t_\mathrm{appl.}$) RCFs to minimise errors due to the continuous post-irradiation development of the RCFs. 

\begin{figure*}[t!]
    \centering
     \pgfplotsset{
compat=1.9, 
}
\tikzset{
  mynode/.style={
  , align=center
  , execute at begin node=\setlength{\baselineskip}{1em}
  }
}

\begin{tikzpicture}

\def\Hseparation{2cm}
\def\Vseparation{1.2cm}

\begin{axis}[
    name=dT_calib, 
    anchor=south west,
    height=0.4\linewidth, 
    width=0.5\linewidth,
    ylabel={$\mathit{nOD}$},
    xlabel={$D_\mathrm{calib.}$~/~Gy},
    grid=both,
    xmin=0, xmax=20,
    ymin=0, ymax=1.4,
    yticklabel style={
      /pgf/number format/fixed,
      /pgf/number format/fixed zerofill,
      /pgf/number format/precision=2
    },
    ytick={0,0.25,0.5,0.75,1,1.25},
    legend style={font=\footnotesize, 
        legend pos =north west},
    ]

    \addplot+[only marks,mark=*,mark size=1.5, mark options={Brown4}, 
        ]
            table [x = dose, y = red,  col sep=comma,restrict x to domain=0:20] {plots/raw_data/calibration/EBT3_08252101_19-10-2022_eRT6_nOD_calibdata.csv}; \label{plot:EBT3_calib1_red_raw}

    \addplot [dashed, color=Brown4,  forget plot
      ]  table [x = fit_doses, y = fit_red, col sep=comma,restrict x to domain=0:20] {plots/raw_data/calibration/EBT3_08252101_19-10-2022_eRT6_nOD_calibdata.csv}; \label{plot:EBT3_calib1_red_fit}
    
    \addplot+[only marks,mark=*,mark size=1.5, mark options={Chartreuse4}, forget plot
            ]  
            table [x = dose, y = green, col sep=comma, restrict x to domain=0:20] {plots/raw_data/calibration/EBT3_08252101_19-10-2022_eRT6_nOD_calibdata.csv}; \label{plot:EBT3_calib1_green_raw}
    
    \addplot [dashed, color=Chartreuse4, forget plot
      ]  table [x = fit_doses, y = fit_green, col sep=comma,restrict x to domain=0:20] {plots/raw_data/calibration/EBT3_08252101_19-10-2022_eRT6_nOD_calibdata.csv}; \label{plot:EBT3_calib1_green_fit}
    
    \addplot+[only marks,mark=*,mark size=1.5, mark options={Blue3}, 
            forget plot]  
            table [x = dose, y = blue,   col sep=comma,restrict x to domain=0:20] {plots/raw_data/calibration/EBT3_08252101_19-10-2022_eRT6_nOD_calibdata.csv}; \label{plot:EBT3_calib1_blue_raw}
    
    \addplot [dashed, color=Blue3, forget plot
      ]  table [x = fit_doses, y = fit_blue, col sep=comma,restrict x to domain=0:20] {plots/raw_data/calibration/EBT3_08252101_19-10-2022_eRT6_nOD_calibdata.csv}; \label{plot:EBT3_calib1_blue_fit}

    \addplot+[only marks,mark=diamond*, mark options={Brown3}, 
        ]
            table [x = dose, y = red,  col sep=comma,restrict x to domain=0:21] {plots/raw_data/calibration/EBT3_08252101_19-10-2022_eRT6_nOD_calibdata_7months.csv}; \label{plot:EBT3_calib3_red_raw}
    
    \addplot [dashed, color=Brown3,  forget plot
      ]  table [x = fit_doses, y = fit_red, col sep=comma,restrict x to domain=0:21] {plots/raw_data/calibration/EBT3_08252101_19-10-2022_eRT6_nOD_calibdata_7months.csv}; \label{plot:EBT3_calib3_red_fit}
    
    \addplot+[only marks,mark=diamond*, mark options={SeaGreen4}, forget plot
            ]  
            table [x = dose, y = green, col sep=comma, restrict x to domain=0:21] {plots/raw_data/calibration/EBT3_08252101_19-10-2022_eRT6_nOD_calibdata_7months.csv}; \label{plot:EBT3_calib3_green_raw}
    
    \addplot [dashed, color=SeaGreen4, forget plot
      ]  table [x = fit_doses, y = fit_green, col sep=comma,restrict x to domain=0:21] {plots/raw_data/calibration/EBT3_08252101_19-10-2022_eRT6_nOD_calibdata_7months.csv}; \label{plot:EBT3_calib_3_green_fit}
    
    \addplot+[only marks,mark=diamond*, mark options={RoyalBlue3}, forget plot
            ]  
            table [x = dose, y = blue,   col sep=comma,restrict x to domain=0:21] {plots/raw_data/calibration/EBT3_08252101_19-10-2022_eRT6_nOD_calibdata_7months.csv}; \label{plot:EBT3_calib3_blue_raw}
    
    \addplot [dashed, color=RoyalBlue3, forget plot
      ]  table [x = fit_doses, y = fit_blue, col sep=comma,restrict x to domain=0:21] {plots/raw_data/calibration/EBT3_08252101_19-10-2022_eRT6_nOD_calibdata_7months.csv}; \label{plot:EBT3_calib3_blue_fit}
    
\end{axis}

\begin{axis}[
    name=drift, 
    at={($(dT_calib.south east) + (\Hseparation,0)$)},
    anchor=south west,
    height=0.4\linewidth, 
    width=0.5\linewidth,
    xlabel={$D_\mathrm{IC}$~/~Gy},
    ylabel=$\Delta D_\mathrm{RCF-IC}/D_\mathrm{IC}$ / \%,
    grid=both,
    legend style={font=\footnotesize, 
        legend pos =north east},
    ]

    \addplot+[dashed, color=Purple4, mark=*,mark size=2, mark options={Purple4}, 
        ]
            table [x = IC, y expr= (\thisrow{calib1}-\thisrow{IC})/\thisrow{IC}*100,  col sep=comma] {plots/raw_data/calibration/calib_drift_multi_nOD.csv}; \label{plot:EBT3_multi_calib1}

    \addplot+[dashed, color=DarkOrchid1,mark=diamond*, mark size=3, mark options={DarkOrchid1}, 
        ]
            table [x = IC, y expr=(\thisrow{calib3}-\thisrow{IC})/\thisrow{IC}*100,  col sep=comma] {plots/raw_data/calibration/calib_drift_multi_nOD.csv}; \label{plot:EBT3_multi_calib3}
    
\end{axis}

\matrix [
            draw,
            fill=white,
            matrix of nodes,
            anchor=north,
            at={($(dT_calib.south east) + (0.5*\Hseparation,-1.7)$)},
        ] {
            $\Delta t_\mathrm{calib.}$ & \qty{24}{\hour} & \qty{233}{\day}\\
            \hline \\
             & \ref{plot:EBT3_calib1_red_raw} \ref{plot:EBT3_calib1_green_raw} \ref{plot:EBT3_calib1_blue_raw}& \ref{plot:EBT3_calib3_red_raw} \ref{plot:EBT3_calib3_green_raw} \ref{plot:EBT3_calib3_blue_raw}\\
             &\ref{plot:EBT3_multi_calib1}  &\ref{plot:EBT3_multi_calib3}\\
        };
\end{tikzpicture}
     \caption{\textbf{Left:} Calibration curves for different $\Delta t_\mathrm{calib.}$ for EBT3 LOT 08252101. \textbf{Right:} Reference RCFs exposed to a known dose $D_\mathrm{calib.}$ and scanned after $\Delta t_\mathrm{appl.}=24~\mathrm{h}$, analysed with two calibration curves of different $\Delta t_\mathrm{calib.}$.}
     \label{fig:EBT3_scans}
\end{figure*}
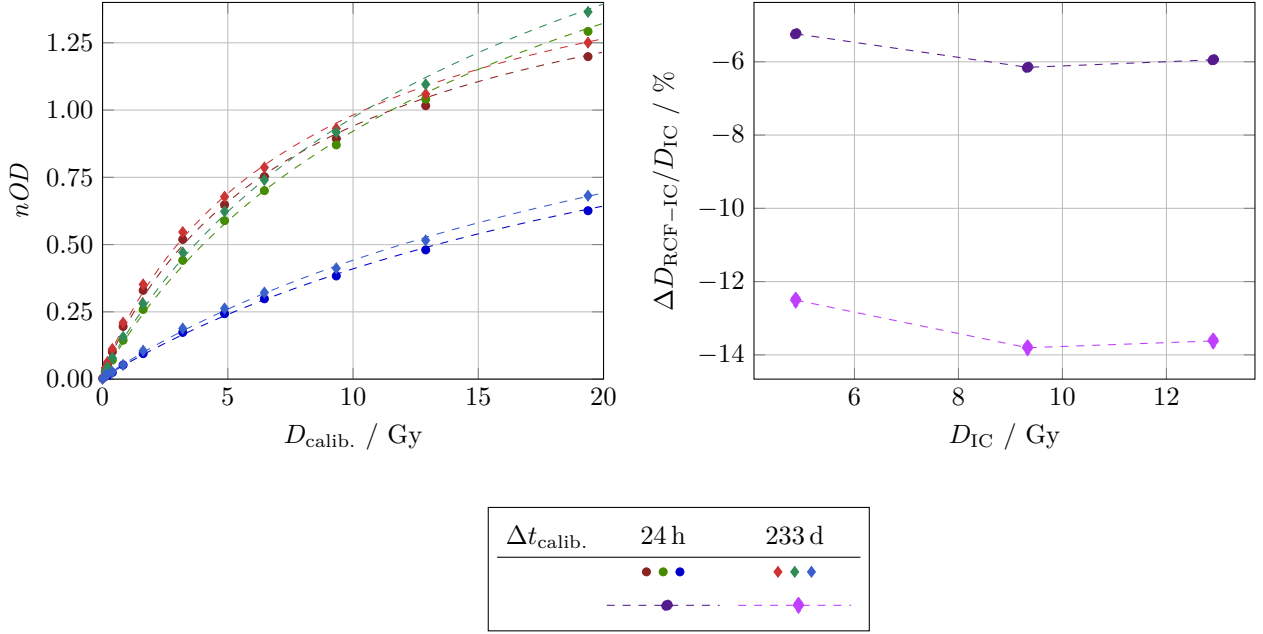

There is a misconception that the background-drift can be compensated for by re-scanning the calibration set simultaneously with scanning the irradiated application RCFs. However, this can introduce a significant error in the calculated dose because of the large time discrepancy between $\Delta t_\mathrm{calib.}$ and $\Delta t_\mathrm{appl.}$. Fig.~\ref{fig:EBT3_scans} illustrates an example when the calibration set was scanned at a $\Delta t_\mathrm{calib.}$ of \qty{24}{\hour} and \qty{233}{\day} post-irradiation.
Using these two calibration curves to process application RCFs of the same lot exposed to known doses at a calibration facility (with an ionisation chamber as reference)\textemdash and scanned at $\Delta t_\mathrm{appl.}\simeq 24~\mathrm{h}$\textemdash we see that an increasing difference between $\Delta t_\mathrm{calib.}$ and $\Delta t_\mathrm{appl.}$ leads to an increasing underestimation of the dose, as per Equation~\ref{eq:dT_criterion}.

For high accuracy, it could be beneficial to update the calibration curve regularly by re-calibrating the batch to account for the natural self-development of the unexposed RCFs subjected to varying environmental conditions. However, we do not have our own calibration setup at CLEAR and so it is a challenge to schedule systematic re-calibrations. 

Typically, a given production lot will be used in CLEAR for up to a year after calibration. To reduce the effect of the ageing of the RCFs, it is advised to ensure stable storage conditions and use nOD as response variable because it effectively accounts for this drift of the background~\cite{transmission_reflection}.

\subsubsection{Summary: The CLEAR Calibration Procedure}
Based on the considerations discussed in section~\ref{sec:rcf_calibration}, the following procedure for RCF calibration has been established at CLEAR:
\begin{enumerate}
  \item Net optical density $\mathit{nOD}$ is used as response variable to account for background drift. 
  \item RCFs are calibrated to a secondary standard ionisation chamber using a low-energy electron linac, that has a sufficiently large and uniform field at a conventional dose-rate. 
  \item 8-12 dose points in geometric progression are used to calibrate for the full dynamic range of a given RCF model. This includes a few (2-3) dose points for calibration verification.
  \item The calibration RCFs are scanned at a time $\Delta t_\mathrm{calib.}\simeq24~\mathrm{h}$. 
  \item The calibration points $\mathit{nOD_x}(D)$ are fitted to the rational function in Eq.~\ref{eq:calib_func}, which expresses the physical nature of RCFs.
\end{enumerate}

\subsection{Image Processing and Analysis}
\label{sec:rcf_processing}
To retrieve dose distributions from irradiated application RCFs, the calibration curves are applied to the digitised images. Dosimetrists use various programs, both commercial and custom, for this analysis. One of the key differences in processing arises from whether the dose is calculated using information from one or multiple of the three (RGB) calibration curves. This section details and compares two approaches\textemdash single-channel and multi-channel processing\textemdash which have both been used at CLEAR.

\subsubsection{Single-Channel RCF Dosimetry}
For accurate dosimetry using a single colour channel, it is important to use the colour channel that exhibits the highest sensitivity:
\begin{equation}
S=\frac{\partial\mathit{nOD}_x}{\partial D}\ .
\label{eq:sensitivity}
\end{equation}

\begin{figure*}[htb!]
    \centering
    \pgfplotsset{compat=1.9}
\tikzset{
  mynode/.style={
  , align=center
  , execute at begin node=\setlength{\baselineskip}{1em}
  }
}


\begin{tikzpicture}

\def\Hseparation{1.5cm}
\def\Vseparation{1.2cm}

\begin{axis}[
    name=ebt3, 
    anchor=south west,
    height=0.4\linewidth, 
    width=0.5\linewidth,
    ylabel={S~/~$\mathrm{Gy}^{-1}$},
    grid=both,
    xmin=0, xmax=20,
    ymin=0, 
    yticklabel style={
      /pgf/number format/fixed,
      /pgf/number format/fixed zerofill,
      /pgf/number format/precision=2
    },
    legend style={font=\footnotesize, 
        legend pos =north west},
    ]
    \addlegendimage{mark=none,black,only marks,mark size=1.5}
    
    \addplot [ color=red, forget plot 
      ]  table [x = fit_doses, y = derivative_red, col sep=comma,restrict x to domain=0:20] {plots/raw_data/calibration/EBT3_08252101_19-10-2022_eRT6_nOD_calibdata.csv}; \label{plot:EBT3_red_derivative}
    \addlegendentry{EBT3 LOT 08252101}
    
    \addplot [ color=green!40!gray,forget plot
      ]  table [x = fit_doses, y = derivative_green, col sep=comma,restrict x to domain=0:20] {plots/raw_data/calibration/EBT3_08252101_19-10-2022_eRT6_nOD_calibdata.csv}; \label{plot:EBT3_green_derivative}

    \addplot [color=blue,forget plot
      ]  table [x = fit_doses, y = derivative_blue, col sep=comma,restrict x to domain=0:20] {plots/raw_data/calibration/EBT3_08252101_19-10-2022_eRT6_nOD_calibdata.csv}; \label{plot:EBT3_blue_derivative}

    \draw [dashed, color=black] (axis cs:0,0.1) -- (axis cs:3.41,0.1)--(axis cs:3.41,0); 
    
    
    
    
    
\end{axis}

\begin{axis}[
    name=ebtxd, 
    at={($(ebt3.south east) + (\Hseparation,0)$)}, anchor=south west,
    height=0.4\linewidth, 
    width=0.5\linewidth,
    grid=both,
    xmin=0, xmax=50,
    ymin=0, 
    yticklabel style={
      /pgf/number format/fixed,
      /pgf/number format/fixed zerofill,
      /pgf/number format/precision=2
    },
    legend style={font=\footnotesize, 
        legend pos =north west},
    ]
    \addlegendimage{mark=*,black,only marks,mark size=1.5}
    
    \addplot [ color=red,  
      ]  table [x = fit_doses, y = derivative_red, col sep=comma,restrict x to domain=0:50] {plots/raw_data/calibration/EBTXD_02232301_19-05-2023_eRT6_nOD_calibdata.csv}; \label{plot:EBTXD_red_derivative}
    \addlegendentry{EBT-XD LOT 02232301}
    
    \addplot [ color=green!70!black,
      ]  table [x = fit_doses, y = derivative_green, col sep=comma,restrict x to domain=0:50] {plots/raw_data/calibration/EBTXD_02232301_19-05-2023_eRT6_nOD_calibdata.csv}; \label{plot:EBTXD_green_derivative}

    \addplot [ color=blue,
      ]  table [x = fit_doses, y = derivative_blue, col sep=comma,restrict x to domain=0:50] {plots/raw_data/calibration/EBTXD_02232301_19-05-2023_eRT6_nOD_calibdata.csv}; \label{plot:EBTXD_blue_derivative}
    \draw [dashed, color=black] (axis cs:0,0.031) -- (axis cs:13.64,0.031)--(axis cs:13.64,0); 
\end{axis}

\begin{axis}[
    name=mdv3, at={($(ebt3.south west) - (0,\Vseparation)$)},anchor=north west, 
    height=0.4\linewidth, 
    width=0.5\linewidth,
    ylabel={S~/~$\mathrm{Gy}^{-1}$},
    xlabel={Dose~/~Gy},
    grid=both,
    xmin=0, xmax=100,
    ymin=0, 
    yticklabel style={
      /pgf/number format/fixed,
      /pgf/number format/fixed zerofill,
      /pgf/number format/precision=2
    },
    legend style={font=\footnotesize, 
        legend pos =north west},
    ]
    \addlegendimage{mark=*,black,only marks,mark size=1.5}

    \addplot [color=red,  
      ]  table [x = fit_doses, y = derivative_red, col sep=comma,restrict x to domain=0:100] {plots/raw_data/calibration/MDV3_10272002_19-05-2023_eRT6_nOD_calibdata.csv}; \label{plot:MDV3_red_derivative}
    \addlegendentry{MD-V3 LOT 10272002}
    
    \addplot [ color=green!40!gray,
      ]  table [x = fit_doses, y = derivative_green, col sep=comma,restrict x to domain=0:100] {plots/raw_data/calibration/MDV3_10272002_19-05-2023_eRT6_nOD_calibdata.csv}; \label{plot:MDV3_green_derivative}

    \addplot [color=blue,
      ]  table [x = fit_doses, y = derivative_blue, col sep=comma,restrict x to domain=0:100] {plots/raw_data/calibration/MDV3_10272002_19-05-2023_eRT6_nOD_calibdata.csv}; \label{plot:MDV3_blue_derivative}
    \draw [dashed, color=black] (axis cs:0,0.0105) -- (axis cs:38.79,0.0105)--(axis cs:38.79,0); 
\end{axis}

\begin{axis}[
    name=hdv2, at={($(ebtxd.south west) - (0,\Vseparation)$)},anchor=north west,
    height=0.4\linewidth, 
    width=0.5\linewidth,
    xlabel={Dose~/~Gy},
    grid=both,
    xmin=0, xmax=1000,
    ymin=0, 
    yticklabel style={
      /pgf/number format/fixed,
      /pgf/number format/fixed zerofill,
      /pgf/number format/precision=2
    },
    legend style={font=\footnotesize, 
        legend pos =north west},
    ]
    \addlegendimage{mark=*,black,only marks,mark size=1.5}
    
    \addplot [ color=red,  
      ]  table [x = fit_doses, y = derivative_red, col sep=comma,restrict x to domain=0:1000] {plots/raw_data/calibration/HDV2_02042104_11-03-2022_eRT6_nOD_calibdata.csv}; \label{plot:HDV2_red_derivative}
    \addlegendentry{HD-V2 LOT 02042104}

    \addplot [ color=green!40!gray,
      ]  table [x = fit_doses, y = derivative_green, col sep=comma,restrict x to domain=0:1000] {plots/raw_data/calibration/HDV2_02042104_11-03-2022_eRT6_nOD_calibdata.csv}; \label{plot:HDV2_green_derivative}
        
    \addplot [color=blue,
      ]  table [x = fit_doses, y = derivative_blue, col sep=comma,restrict x to domain=0:1000] {plots/raw_data/calibration/HDV2_02042104_11-03-2022_eRT6_nOD_calibdata.csv}; \label{plot:HDV2_blue_derivative}

    \draw [dashed, color=black] (axis cs:0,0.0005) -- (axis cs:989.9,0.0005)--(axis cs:989.9,0); 
    
\end{axis}

\end{tikzpicture}


    \caption{Sensitvity of the different calibration functions in Fig.~\ref{fig:calib_curves}.}
    \label{fig:derivatives}
\end{figure*}
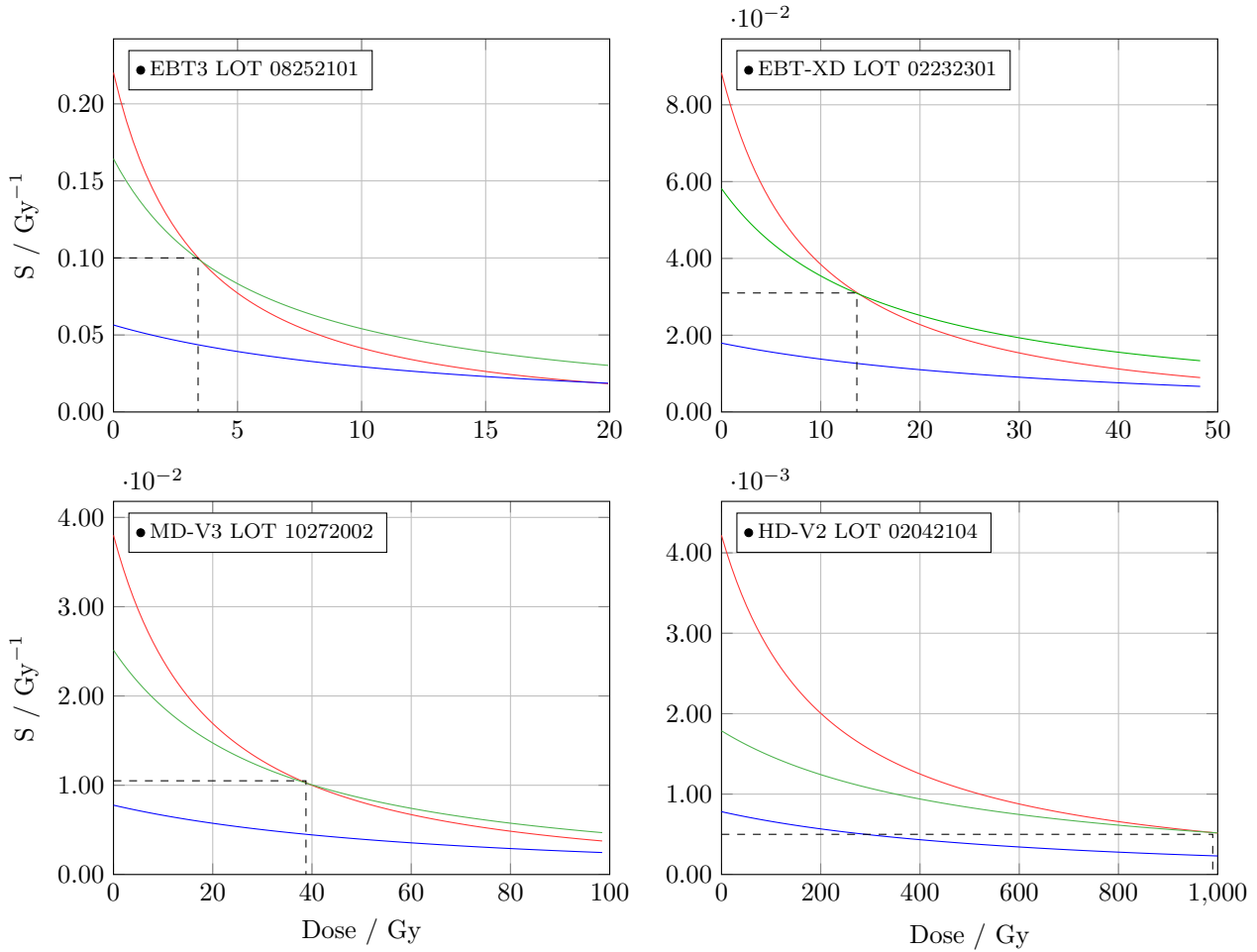

Literature reviews indicate that the red channel is the most commonly used channel for single channel dosimetry on the basis that Gafchromic\texttrademark~RCFs have maximum absorption in the red band ($\sim 633~\mathrm{nm}$). However, as shown in Fig.~\ref{fig:calib_curves} in section~\ref{sec:rcf_calibration}, the red channel is not necessarily the most sensitive channel for all dose ranges and RCF models. The derivatives of the calibration functions are displayed in Fig.~\ref{fig:derivatives} and shows, for example, that beyond 3\textendash 4 Gy, the green channel is more sensitive than the red for EBT3. It is also evident why the blue channel is hardly used for single-channel RCF dosimetry, although it should be noted that it has been shown that it can indeed be useful to extend the dynamic range of the RCFs~\cite{devic_range}.

When using a single colour channel $x$, the dose is calculated directly from the inverse of the calibration function in Eq.~\ref{eq:calib_func} as
\begin{equation}
D = \frac{a_x - c_x \mathrm{e}^{-\mathit{nOD}_x}}{\mathrm{e}^{-\mathit{nOD}_x} - b_x}\ .
\label{eq:dose_func}
\end{equation}
The related uncertainty may be estimated as:

\begin{strip}
    \begin{equation}
    \sigma_{D}=\sqrt{\left(\frac{\partial D}{\partial \mathit{nOD}}\right)^2\sigma_{\mathit{nOD}}^2+\left(\frac{\partial D}{\partial a_x}\right)^2\sigma_{a_x}^2+\left(\frac{\partial D}{\partial b_x}\right)^2\sigma_{b_x}^2+\left(\frac{\partial D}{\partial c_x}\right)^2\sigma_{c_x}^2}, 
\label{eq:dose_uncertainty}
\end{equation}
\end{strip}

where $\sigma_{\mathit{nOD}}$ is calculated from Eq.~\ref{eq:sigma_nod}, and $\sigma_{a_x}$, $\sigma_{b_x}$ and $\sigma_{c_x}$ are the uncertainties in the fitting parameters obtained when establishing the calibration curve. 

A single-channel processing method was used at CLEAR between 2019 and 2024, employing custom-written code to calculate dose distributions for small Gaussian beams. However, when using a single colour channel to determine the dose on the RCF, the entire response is converted to dose. Due to the many described uncertainties and fluctuations originating from RCF handling and scanner response, this may lead to potentially large errors in dose estimation, because some artefacts yield larger response in specific channels. This was the motivation for transitioning to multi-channel dosimetry, which weights the responses from the different channels and thereby minimises the influence of non-dose-dependent artifacts.

\subsubsection{Multi-Channel RCF Dosimetry}
Some facilities have access to commercial tools such as FilmQAPro (Ashland) and radiochromic.com for RCF processing. These  programs have the capability for multi-channel processing that uses information from all three colour channels for dose evaluation. Multi-channel processing exploits information from all three colour channels to separate the RCF response into a dose-dependent response and a dose-independent perturbation~\cite{rcf_composition}. By estimating the perturbation which minimises the difference between doses calculated from each individual colour channel, the noise from the dose-distribution is reduced. 

The multi-channel method can mitigate issues such as variations in thickness across the active layer, and scanner artifacts\textemdash including LRA, noise and, to some extent, dust\textemdash by optimising the dose value to the most probable value using information from all three calibration curves. The multi-channel method was first proposed by Micke et al.~\cite{micke_multi}, and Mayer et al. later proposed an easily-implementable solution~\cite{mayer_multi}. This solution expresses the dose as a Taylor expansion with a perturbation, and minimises the cost function\textemdash which represents the difference between the true (common) dose $D$ and the channel-dependent responses, which includes both the dose and the perturbation. 

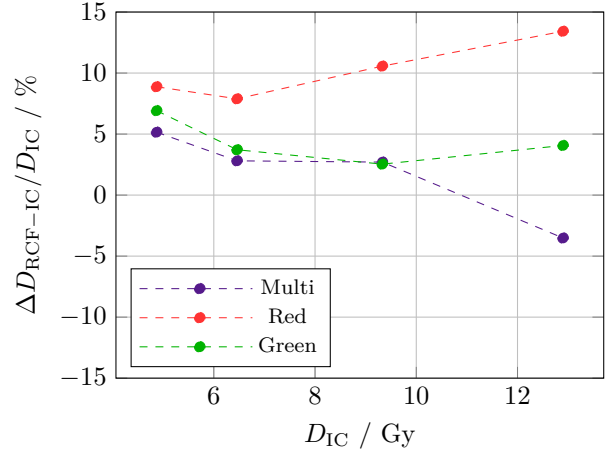
\begin{figure}[htb!]
    \centering
    \begin{tikzpicture}

\begin{axis}[
    name=drift, 
    anchor=south west,
    height=0.8\linewidth, 
    width=\linewidth,
    xlabel={$D_\mathrm{IC}$~/~Gy},
    ylabel=$\Delta D_\mathrm{RCF-IC}/D_\mathrm{IC}$ / \%,
    grid=both,
    ymin=-15, ymax=15,
    ytick={-15,-10,-5,0,5,10,15},
    legend style={font=\footnotesize, 
        legend pos =south west},
    ]

    \addplot+[dashed, color=Purple4, mark=*,mark size=2, mark options={Purple4}, 
        ]
            table [x = IC, y expr= (\thisrow{CLEAR dose}-\thisrow{IC})/\thisrow{IC}*100,  col sep=comma] {plots/raw_data/CLEARvsCHUV.csv}; \label{plot:multi_channel}
    \addlegendentry{Multi}

    \addplot+[dashed, color=red,mark=*, mark size=2, mark options={red}, 
        ]
            table [x = IC, y expr=(\thisrow{CLEAR dose}-\thisrow{IC})/\thisrow{IC}*100,  col sep=comma] {plots/raw_data/CLEARvsCHUV_red.csv}; \label{plot:red_channel}
    \addlegendentry{Red}
    \addplot+[dashed, color=green!70!black,mark=*, mark size=2, mark options={green!70!black}, 
        ]
            table [x = IC, y expr=(\thisrow{CLEAR dose}-\thisrow{IC})/\thisrow{IC}*100,  col sep=comma] {plots/raw_data/CLEARvsCHUV_green.csv}; \label{plot:green_channel}
   \addlegendentry{Green}
    
\end{axis}
\end{tikzpicture}
    \caption{The dose errors of EBT3 RCFs exposed to a known dose $D_\mathrm{IC}$ at a calibration facility and processed using single-channel processing and multi-channel processing.}
    \label{fig:channel_comparison}
\end{figure}

As of 2025, a custom-code for multi-channel processing has been implemented at CLEAR, whereby the dose for each pixel of the application RCF is evaluated as
\begin{equation}
D=\frac{\langle D \rangle -\mathit{RS}\cdot \frac{\sum_x D_x\cdot D'_x}{\sum_x D'_x}}{1-\mathit{RS}}\ .
\end{equation}
Here, $\langle D \rangle$ is the average dose estimated using all colour channels:

\begin{equation}
    \langle D \rangle=\frac{1}{3}\sum_x D_x\ ,
\end{equation}

and $D'_x$ is the derivative of the dose function w.r.t. the response variable $\mathit{nOD}_x$
\begin{equation}
D'_x=\frac{\partial D(\mathit{nOD}_x)}{\partial \mathit{nOD}_x}=\mathrm{e}^{\mathit{nOD}_x}\cdot\frac{a_x-b_xc_x}{(b_x\mathrm{e}^{\mathit{nOD}_x}-1)^2}\ ,
\end{equation}
while $\mathit{RS}$ is the relative slope of the colour channels
\begin{equation}
\mathit{RS}=\frac{1}{3}\frac{(\sum_x D'_x)^2}{\sum_x {D'}_x^2}\ .
\end{equation}
The corresponding uncertainty is then calculated by further propagating the single channel dose uncertainty in Eq.~\ref{eq:dose_uncertainty}.
The improved accuracy of using multi-channel processing over single-channel is shown in Fig.~\ref{fig:channel_comparison}. We see that the relative dose error resulting from multi-channel dosimetry is within $\pm\SI{5}{\percent}$, which is lower than the error associated with single-channel dosimetry using the green and red channels.
\subsubsection{Summary: The CLEAR RCF Processing Procedure}
Based on the considerations described in section~\ref{sec:rcf_processing}, the following procedure is followed for RCF processing at CLEAR:
\begin{enumerate}
  \item Application RCFs are scanned after $\Delta t_\mathrm{appl.}\simeq\Delta t_\mathrm{calib.}\simeq24~\mathrm{h}$. 
  \item Between 2019-2024, the response was converted to dose using the single-channel method for the colour channel of highest sensitivity in the dose-regime in question.
  \item Since 2025, the multi-channel method has been implemented, and is used for noise reduction and higher accuracy.
\end{enumerate}

\section{Comparative Measurements with Passive Dosimeters}
\label{sec:benchmarking}

To evaluate both the accuracy of our RCF dosimetry for typical CLEAR experiments and the robustness of the described RCF dosimetry protocol, a series of experiments were conducted comparing our RCF dosimetry with three other passive dosimeters: alanine dosimeters (ADs), radio-photoluminescence dosimeters (RPLDs), and dosimetry phantoms (DPs). These comparative measurements were performed using a geometry similar to that of a typical irradiation at CLEAR—where the ADs, RPLDs, and DPs were positioned comparably to standard samples relative to the RCFs.

\subsection{Irradiation Setup at CLEAR}
CLEAR has a Cartesian robot with capability to position custom 3D-printed sample holders into and out of the electron beam without manual intervention~\cite{pierre22}. The robot offers significant advantages in typical CLEAR experiments by ensuring reproducible sample positioning and allowing multiple samples to be irradiated in a relatively short time—without the need to wait for the required 30-minute cool-down period to access the machine.

\begin{figure*}[htb!]
   \centering
   \includegraphics{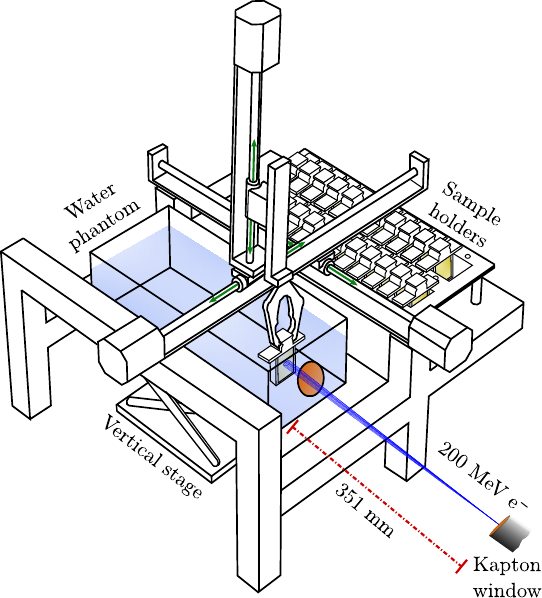}
   \caption{A drawing of the robot holding a sample holder inside the water phantom. The path of the beam is indicated in blue. Illustration adapted from~\cite{rieker_nima}.}
   \label{fig:robot_sketch}
\end{figure*}

An illustration of the irradiation setup including the robot is shown in Fig.~\ref{fig:robot_sketch}. The robot employs a sample grabber to pick up sample holders from a storage area outside of the beam and place them in front of the beam. The sample storage area has a capacity of up to 32 sample holders and optional temperature control. Additionally, a water phantom mounted on a vertical stage is positioned within the robot's boundaries along the beam path, enabling sample irradiations in water. 

Typical CLEAR irradiation experiments are undertaken inside the water phantom. The irradiation depth depends on the beam conditions, which are not constant in CLEAR. Most irradiations at CLEAR have so far been performed with a Gaussian beam, and the depth of irradiation is typically determined by the beam-size requirement of the experiment. 

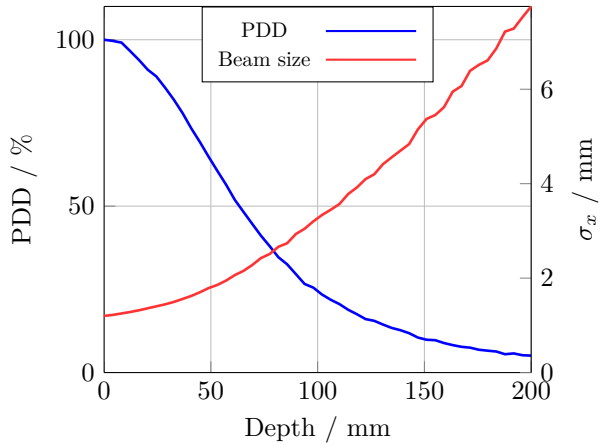
\begin{figure}[htb!]
    \centering
\pgfplotsset{compat=1.9}

\begin{tikzpicture}
\begin{axis}[
width=0.9\linewidth,
height=0.8\linewidth,
name=dmax,
axis y line*=left,
  xlabel={Depth~/~mm},
  ylabel={PDD~/~\%},
  grid=both,
  xmin=0, xmax=200,
  ymin=0, 
]

\addplot+ [color=blue, line width=1, 
  ]  table [x = depths, y = PDD, col sep=comma] {plots/raw_data/sim_data.csv}; \label{plot:PDD}
\end{axis}

\begin{axis}[
width=0.9\linewidth,
height=0.8\linewidth,
    at={(dmax.south west)},
    anchor=south west,
    enlargelimits=false,
    axis y line*=right,
    axis x line=none,
    ylabel={$\sigma_x$~/~mm},
    xmin=0, xmax=200,
    ymin=0,
      ]
      \addplot+ [color=red, line width=1, 
  ]  table [x = depths, y = sigY, col sep=comma] {plots/raw_data/sim_data.csv}; \label{plot:sigX}
\end{axis}

\matrix [
            draw,
            fill=white,
            matrix of nodes,
            nodes={scale=0.8, mark options={scale=0.5}},
            anchor=north,
            ] at (dmax.north) {
                PDD & \ref{plot:PDD}\\
                Beam size  & \ref{plot:sigX} \\
            };

\end{tikzpicture}
     \caption{Typical percentage-depth-dose (PDD) curve for 200 MeV electrons in water, as simulated with TOPAS.}
     \label{fig:CLEAR_pdd}
\end{figure}

Fig.~\ref{fig:CLEAR_pdd} shows a TOPAS simulation of a typical evolution of peak dose and beam-size throughout the water phantom for a Gaussian beam at CLEAR~\cite{topas}. The initial Twiss parameters of the beam affect the steepness of the curves, but the overall evolution is as expected for a Gaussian VHEE beam as opposed to the flatter build-up for uniform beams as demonstrated by Böhlen \textit{et al.}~\cite{Bohlen2021CharacteristicsTargets}. Similar simulations have previously been confirmed experimentally to be representative of the beam at CLEAR~\cite{Rieker2023BeamFacility}.

For the comparative measurements between the RPLDs, DPs and RCFs, different mean dose rates ($\langle \dot{D}\rangle$) were used to study the relative agreement at both CONV and UHDR. CONV irradiation conditions at the CLEAR facility use one bunch per pulse with a bunch charge in the order of \SI{100}{\pico\coulomb} and a bunch length of approximately 1\textendash\SI{5}{\pico\second}, delivered at a pulse repetition rate of \SI{0.833}{\hertz}. On the other hand, UHDR irradiation conditions deliver a single electron pulse that consists of a train of bunches spaced at a frequency of \SI{1.5}{\giga\hertz}. The charge per pulse (and hence dose per pulse) is determined by the number of bunches within the train and hence the pulse width (up to $\sim$\SI{50}{\nano\second}).
\subsection{Comparison with Alanine Dosimeters}
\label{sec:alanine}
\begin{figure*}[htb!]
     \centering
     \input{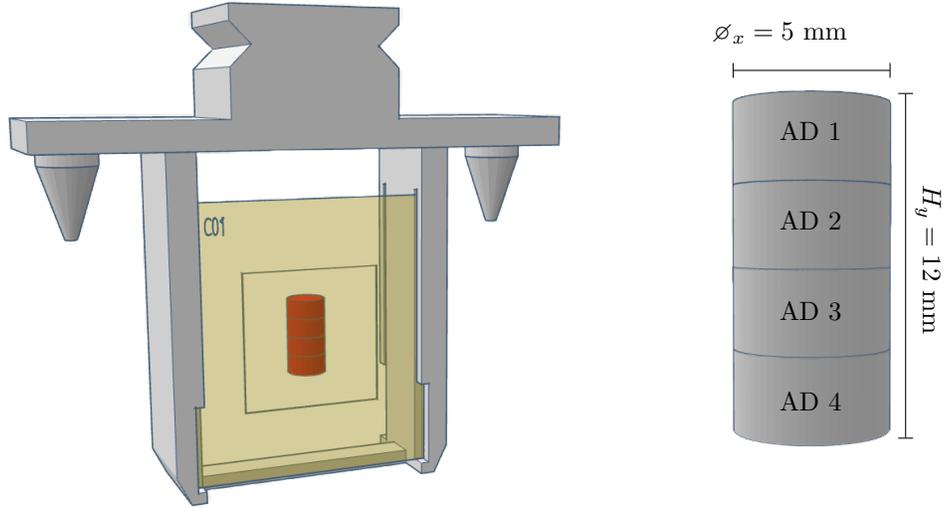}
    \caption{\textbf{Left:} The robot holder with a stack of four alanine pellets (indicated in red) packed in plastic and attached on the back of an RCF. \textbf{Right:} The stack of ADs with total dimensions.}
    \label{fig:alanine_setup}
\end{figure*}

The aim of the study was to compare the overall agreement between the RCF dosimetry at CLEAR and alanine dosimeters (ADs) from PTB, but also to see if the agreement remains the same for different energies. The experimental measurements compared the dose response of RCFs and ADs and was undertaken at CLEAR in collaboration with Physikalisch-Technische Bundesanstalt (PTB) in Germany. Stacks of four ADs (each measuring \SI{5}{\milli\metre} in diameter and \SI{3}{\milli\metre} in height), shrink-wrapped in polythene foil were positioned on the downstream side of an EBT-XD RCF and mounted in a robot holder as shown in Fig. \ref{fig:alanine_setup}. 

 Doses of $10$, $15$, and \SI{20}{\gray} were targeted each at different beam energies of $50$, $100$, $150$, and \SI{200}{\mega\electronvolt}. delivered using the CONV irradiation conditions. For each measurement, the beam size was kept constant at $\sim\sigma_{x,y} =$ \SI{5.5}{\milli\metre} to ensure that all samples were exposed to a similar radiation field. This required adjustment of the irradiation depth at the different energies.

Both the single green channel, as the overall most sensitive channel for EBT-XD in the range 10\textendash 20 Gy, and multi-channel processing methods were used to evaluate the dose response of the RCFs. The mean RCF doses and standard deviations were obtained from the ROI corresponding to the positions of the two central ADs (AD2 and AD3) to optimise in terms of symmetry and uniformity. The absorbed doses to the ADs were subsequently determined at PTB by measuring the concentration of free radicals produced by ionising radiation using electron spin resonance (ESR) spectroscopy~\cite{Bourgouin2022Absorbed-dose-to-waterBeams, Anton2014ABIPM, Anton2006UncertaintiesBundesanstalt}. 

\begin{figure*}[htb!]
    \centering
    \pgfplotsset{compat=1.9}
\tikzset{
  mynode/.style={
  , align=center
  , execute at begin node=\setlength{\baselineskip}{1em}
  }
}

\begin{tikzpicture}

\def\Hseparation{2cm}
\def\Vseparation{1.2cm}

\begin{axis}[
    name=alanine, 
    anchor=south west,
    height=0.5\linewidth, 
    width=0.65\linewidth,
    ylabel={Dose~/~Gy},
    xlabel={Holder number},
    grid=both,
    xmin=0.5, xmax=20.5,
    xtick={1,2,...,20},
    xtick pos=bottom,
    ymin=0, ymax=27,
    error bars/y dir=both,
    error bars/y explicit,
    yticklabel style={
        font=\footnotesize, 
    },
    xticklabel style={
        font=\footnotesize, 
    },
    legend style={font=\footnotesize, 
        legend pos =south east},
    legend cell align={left},
    ]

    \fill[gray, opacity=0.1] (axis cs:0,0) rectangle (axis cs:5.5, 27); \node[align=center,anchor=center]at (axis cs:3,25){\footnotesize 50 MeV};

    \node[align=center,anchor=center]at (axis cs:8,25){\footnotesize 100 MeV};

    \fill[gray, opacity=0.1] (axis cs:10.5,0) rectangle (axis cs:15.5, 27); \node[align=center,anchor=center]at (axis cs:13,25){\footnotesize 150 MeV};
    \node[align=center,anchor=center]at (axis cs:18,25){\footnotesize 200 MeV};
    
    \addplot+[ybar, bar width=2.5pt,fill, color=Chartreuse4!50, error bars/error bar style={thick, Chartreuse4, mark size=2pt}, area legend]  
            table [x expr= \thisrow{Holder}-0.25, y = Film Green Dose, y error = Film Green Error, col sep=comma] {plots/raw_data/alanine_barchart.csv}; \label{plot:alanine_bar_green}

    \addplot+[ybar, bar width=2.5pt,fill, color=violet!50, error bars/error bar style={thick, violet, mark size=2pt}, area legend]  
            table [x expr= \thisrow{Holder}-0, y = Film Multi Dose, y error = Film Multi Error, col sep=comma] {plots/raw_data/alanine_barchart.csv}; \label{plot:alanine_bar_multi}

    \addplot+[ybar,bar width=2.5pt,fill, color=black!50, error bars/error bar style={thick, black, mark size=2pt}, area legend]  
            table [x expr= \thisrow{Holder}+0.25, y = Alanine Dose, y error = Alanine Error, col sep=comma] {plots/raw_data/alanine_barchart.csv}; \label{plot:alanine_bar_alanine}

\end{axis}

\begin{axis}[
    name=box, 
    at={($(alanine.south east) + (\Hseparation,0)$)},
    boxplot/draw direction=y,
    anchor=south west,
    height=0.5\linewidth, 
    width=0.35\linewidth,
    xlabel=Energy / MeV,
    ylabel=$\Delta D_{RCF-AD}/D_{AD}$ / \%,
    ymajorgrids=true,
    xmin=0.5, xmax=8.5,
    ymin=-6.5, ymax=4.5,
    xtick={1.5,3.5,5.5,7.5},
    xticklabels={50,100,150,200},
    yticklabel style={
        font=\footnotesize, 
    },
    xticklabel style={
        font=\footnotesize, 
    },
    legend style={font=\footnotesize, 
        legend pos =north west},
    ]
    \fill[gray, opacity=0.2] (axis cs:0.5,-6.5) rectangle (axis cs:2.5, 4.5); 

    \addplot+ [only marks, mark=*,Chartreuse4!50]
    table [x expr=1*\coordindex/\coordindex,y = Film Green 50, col sep=comma] {plots/raw_data/alanine_deviation.csv}; 
    
    \addplot+ [only marks, mark=-, mark size=6, line width=2,Chartreuse4]
    table [x expr=1*\coordindex/\coordindex,y = Film Green 50 avg, col sep=comma] {plots/raw_data/alanine_deviation.csv}; 
    
    \addplot+ [only marks, mark=*, violet!50]
    table [x expr=2*\coordindex/\coordindex, y = Film Multi 50, col sep=comma] {plots/raw_data/alanine_deviation.csv}; 

    \addplot+ [only marks, mark=-, mark size=6, line width=2,violet]
    table [x expr=2*\coordindex/\coordindex, y = Film Multi 50 avg, col sep=comma] {plots/raw_data/alanine_deviation.csv}; 

    \addplot+ [only marks, mark=*,Chartreuse4!50]
    table [x expr=3*\coordindex/\coordindex,y = Film Green 100, col sep=comma] {plots/raw_data/alanine_deviation.csv}; 
    
    \addplot+ [only marks, mark=-, mark size=6, line width=2,Chartreuse4]
    table [x expr=3*\coordindex/\coordindex,y = Film Green 100 avg, col sep=comma] {plots/raw_data/alanine_deviation.csv}; 
    
    \addplot+ [only marks, mark=*, violet!50]
    table [x expr=4*\coordindex/\coordindex, y = Film Multi 100, col sep=comma] {plots/raw_data/alanine_deviation.csv}; 

    \addplot+ [only marks, mark=-, mark size=6, line width=2,violet]
    table [x expr=4*\coordindex/\coordindex, y = Film Multi 100 avg, col sep=comma] {plots/raw_data/alanine_deviation.csv}; 

    \fill[gray, opacity=0.2] (axis cs:4.5,-6.5) rectangle (axis cs:6.5, 4.5); 
    
    \addplot+ [only marks, mark=*,Chartreuse4!50]
    table [x expr=5*\coordindex/\coordindex,y = Film Green 150, col sep=comma] {plots/raw_data/alanine_deviation.csv}; 
    
    \addplot+ [only marks, mark=-, mark size=6, line width=2,Chartreuse4]
    table [x expr=5*\coordindex/\coordindex,y = Film Green 150 avg, col sep=comma] {plots/raw_data/alanine_deviation.csv}; 
    
    \addplot+ [only marks, mark=*, violet!50]
    table [x expr=6*\coordindex/\coordindex, y = Film Multi 150, col sep=comma] {plots/raw_data/alanine_deviation.csv}; 

    \addplot+ [only marks, mark=-, mark size=6, line width=2,violet]
    table [x expr=6*\coordindex/\coordindex, y = Film Multi 150 avg, col sep=comma] {plots/raw_data/alanine_deviation.csv}; 

    
    \addplot+ [only marks, mark=*,Chartreuse4!50]
    table [x expr=7*\coordindex/\coordindex,y = Film Green 200, col sep=comma] {plots/raw_data/alanine_deviation.csv}; 
    
    \addplot+ [only marks, mark=-, mark size=6, line width=2,Chartreuse4]
    table [x expr=7*\coordindex/\coordindex,y = Film Green 200 avg, col sep=comma] {plots/raw_data/alanine_deviation.csv}; 
    
    \addplot+ [only marks, mark=*, violet!50]
    table [x expr=8*\coordindex/\coordindex, y = Film Multi 200, col sep=comma] {plots/raw_data/alanine_deviation.csv}; 

    \addplot+ [only marks, mark=-, mark size=6, line width=2,violet]
    table [x expr=8*\coordindex/\coordindex, y = Film Multi 200 avg, col sep=comma] {plots/raw_data/alanine_deviation.csv}; 
\end{axis}

\matrix [
            draw,
            fill=white,
            matrix of nodes,
            nodes={ minimum size=10pt, minimum width=40pt, font=\footnotesize},
            minimum width=0.38\linewidth,
            anchor=north,
            at={($(alanine.south east) + (0.5*\Hseparation,-1.5)$)},
        ] {
             \ref{plot:alanine_bar_green} $\mathrm{RCF_{green}}$ & \ref{plot:alanine_bar_multi} $\mathrm{RCF_{multi}}$ & \ref{plot:alanine_bar_alanine} AD\\
        };

\end{tikzpicture}
    \caption{\textbf{Left:} Bar chart of dose measurements from each irradiation holder of the alanine pellets and EBT-XD RCFs analysed for both single green channel and with the multi-channel method. The AD doses are the averages of the two central pellets (AD2 and AD3) and compared for 50, 100, 150 and 200 MeV at CLEAR. \textbf{Right:} The deviations between RCF and AD measurements for each beam energy.}
    \label{fig:alaninedoses}
\end{figure*}
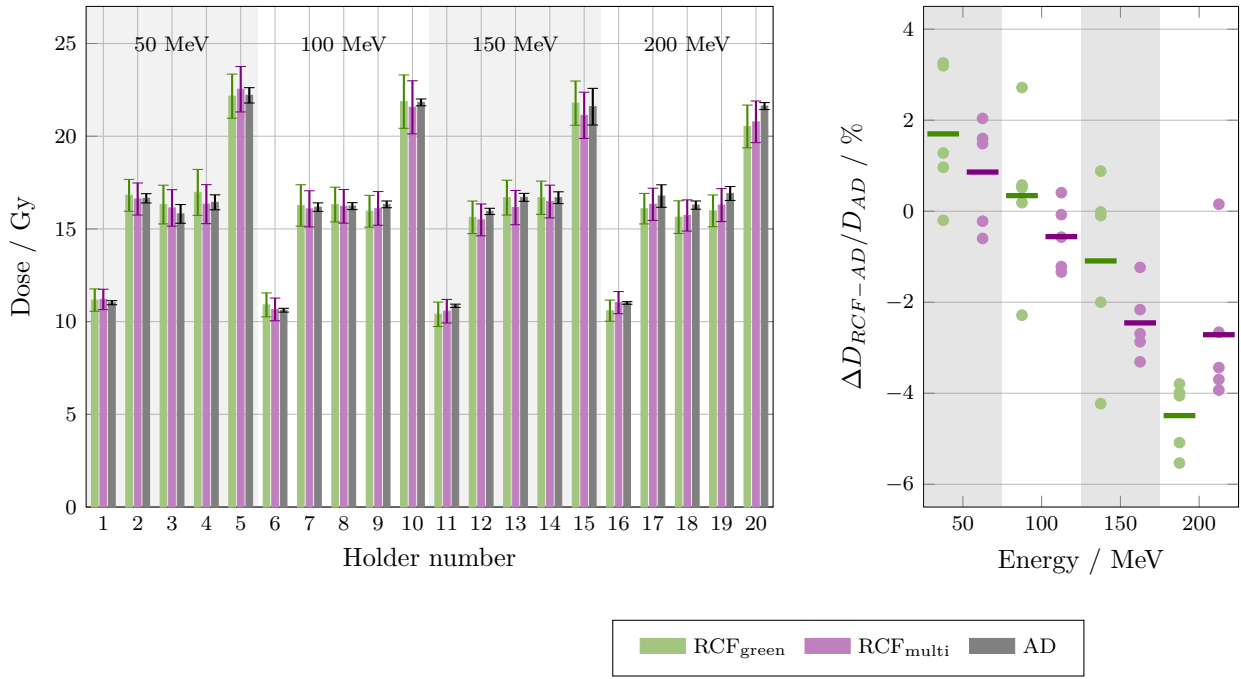

The dose responses of the RCFs and ADs are compared in Fig.~\ref{fig:alaninedoses}. The left-hand plot shows that the dose values obtained from both the single green channel and multi-channel processing of the RCFs agree with the AD value to within a single standard deviation from each dosimeter. 

The mean standard deviation of the RCF measurements (\SI{5.8}{\percent}) is larger than for the AD measurements (\SI{1.7}{\percent}) for multiple reasons.  The dominant contribution is because the small Gaussian beam ($\sigma_{x,y} = \SI{5.5}{\milli\metre}$) used in CLEAR is inhomogeneous across the area occupied by the ADs ($\diameter_x = \SI{5}{\milli\metre}$, $H_y = \SI{3}{\milli\metre}$) whereas the AD calibration refers to a homogeneous reference field of $10\times\SI{10}{\centi\metre\squared}$. Small differences in lateral positioning of the AD with regard to the beam axis in small fields can lead to large deviations in the dose that are difficult to assess. Furthermore, the influence of the inhomogeneous distribution of the free radicals on the ESR signal is not easily estimated.

Another uncertainty arises due to the deviations from reference conditions. The ADs were positioned in a small Gaussian field at considerable depths in water (\SIrange{62}{180}{\milli\metre}), while the AD calibration refers to Co-60 irradiations under reference conditions\textemdash a homogeneous field of $10\times\SI{10}{\centi\metre\squared}$ at \SI{50}{\milli\metre} depth in water\textemdash accounting for a radiation quality factor (1.012) that corrects for the AD response to electron radiation under reference conditions (homogeneous field of $10\times\SI{10}{\centi\metre\squared}$ at reference depths of \SIrange{10}{60}{\milli\metre}). For small electron fields, such as those at CLEAR, a reference depth is not even defined.

Lastly, the validity of the AD calibration is only verified for 6\textendash\SI{20}{\mega\electronvolt} electrons~\cite{Voros2012RelativeStandard}. Although no significant energy dependence has been observed in the range 15\textendash\SI{50}{\mega\electronvolt} by comparison with RCF and IC measurements~\cite{Kokurewicz2021AnRadiotherapy}, the energy range (50\textendash\SI{200}{\mega\electronvolt}) used for irradiation at CLEAR could also be a source of uncertainty.

The right-hand plot in Fig.~\ref{fig:alaninedoses} shows that both the single green channel and multi-channel RCF processing methods yield average absolute relative deviations of less than \SI{5}{\percent} for all beam energies compared to the AD measurements. Overall, the multi-channel method seems to perform better relative to the AD values.

\subsection{Comparison with Radio-photoluminescence Dosimeters}
\label{sec:rpl}
The dose response of RCFs used in CLEAR was also compared to radio-photoluminescence dosimeter (RPLD) readings, with the additional aim of investigating their relative agreement between CONV and UHDR dose-rates. RPLDs are passive dosimeters made from silver doped phosphate glass. They offer advantages of being robust, exhibiting minimal fading effects, and allowing for repeated readouts without signal loss. Under radiation exposure, electron-hole pairs generated within the glass are trapped and give rise to two types of optical centres: luminescence centres (RPL centres), and colour centres. The principle of operation for these passive dosimeters is the proportionality between number of radiation-induced luminescence centres, and the deposited dose. Under UV light exposure, the RPL centres emit luminescence light that can be measured for the estimation of absorbed dose in the glass dosimeter~\cite{rpl_yamamoto}. This dosimetry technology is largely used in medical applications and for environmental monitoring but, at CERN, their sensitive range goes beyond their traditional application, reaching the \unit{\mega\gray}-range~\cite{rpl_aguiar}. In a previous study, we compared the agreement between RPLDs and HD-V2 RCFs in air in the dose-range 30\textendash\SI{300}{\gray}, where an agreement within \SI{10}{\percent} was found~\cite{ipac22}. However, the question remained whether a better agreement could be achieved in water for clinical doses and whether there was a clear dose-rate dependency.

\begin{figure*}[htb!]
    \centering
    \input{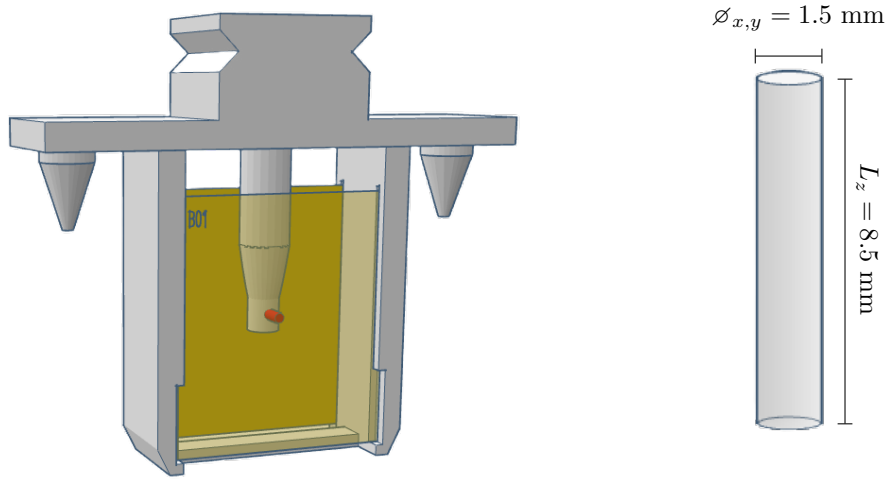}
    \caption{\textbf{Left:} the robot holder with an RPLD (indicated in red) positioned between two RCFs. \textbf{Right:} the RPLD glass cylinder with dimensions. }
    \label{fig:rpl_setup}
\end{figure*}

While the ADs were fixed directly downstream of an RCF, the cylindrical RPLDs were, due to their larger longitudinal ($z$) size,  positioned between two RCFs as shown in Fig.~\ref{fig:rpl_setup}. 16 RPLDs combined with both EBT-XD and MD-V3 RCFs were irradiated at both UHDR and CONV, targeting doses of 10, 15 and 20 Gy. These doses are typical for medical application experiments at CLEAR.

\label{subsubsec:rpl_setup}

\begin{figure*}[htb!]
    \centering
    \pgfplotsset{compat=1.9}
\tikzset{
  mynode/.style={
  , align=center
  , execute at begin node=\setlength{\baselineskip}{1em}
  }
}

\begin{tikzpicture}

\def\Hseparation{2cm}
\def\Vseparation{1.2cm}

\begin{axis}[
    name=rpl,
    anchor=south west,
    height=0.5\linewidth, 
    width=0.7\linewidth,
    ylabel={Dose~/~Gy},
    xlabel={Holder number},
    grid=both,
    xmin=0.5, xmax=16.5,
    xtick={1,2,...,16},
    xtick pos=bottom,
    ymin=0, ymax=30,
    error bars/y dir=both,
    error bars/y explicit,
    yticklabel style={
        font=\footnotesize, 
    },
    xticklabel style={
        font=\footnotesize, 
    },
    legend style={font=\footnotesize, 
        legend pos =south east},
    legend cell align={left},
    ]

    \fill[gray, opacity=0.1] (axis cs:0,0) rectangle (axis cs:8.5, 30); \node[align=center,anchor=center]at (axis cs:4.5,28){\footnotesize EBT-XD};

    \node[align=center,anchor=center]at (axis cs:12.5,28){\footnotesize MD-V3};
    
    \addplot+[ybar, bar width=4pt,fill, color=Chartreuse4!50, error bars/error bar style={thick, Chartreuse4, mark size=2.5pt}, area legend]  
            table [x expr= \thisrow{Holder}-0.25, y = Film Green Avg, y error = Film Green Avg Err, col sep=comma] {plots/raw_data/rpl_barchart.csv}; \label{plot:rpl_bar_green}

    \addplot+[only marks, color=Chartreuse4, mark=triangle, forget plot, error bars/.cd, y dir=both, y explicit]
    table [x expr= \thisrow{Holder}-0.25, y = Film Green US, y error = Film Green US Err, col sep=comma] {plots/raw_data/rpl_barchart.csv}; \label{plot:rpl_US_green}
    
    \addplot+[only marks, color=Chartreuse4, mark=triangle*, mark options={rotate=180}, forget plot, error bars/.cd, y dir=both, y explicit]
    table [x expr= \thisrow{Holder}-0.25, y = Film Green DS, y error = Film Green DS Err, col sep=comma] {plots/raw_data/rpl_barchart.csv}; \label{plot:rpl_DS_green}

    \addplot+[ybar, bar width=4pt,fill, color=violet!50, error bars/error bar style={violet}, area legend]  
            table [x expr= \thisrow{Holder}-0, y = Film Multi Avg, y error = Film Multi Avg Err, col sep=comma] {plots/raw_data/rpl_barchart.csv}; \label{plot:rpl_bar_multi}

     \addplot+[xshift=-0pt, only marks, color=violet, mark=triangle, forget plot, error bars/.cd, y dir=both, y explicit]
    table [x expr= \thisrow{Holder}-0, y = Film Multi US, y error = Film Multi US Err, col sep=comma] {plots/raw_data/rpl_barchart.csv}; \label{plot:rpl_US_multi}
    \addplot+[xshift=-0pt,only marks, color=violet, mark=triangle*, mark options={rotate=180}, forget plot, error bars/.cd, y dir=both, y explicit]
    table [x expr= \thisrow{Holder}-0, y = Film Multi DS, y error = Film Multi DS Err, col sep=comma] {plots/raw_data/rpl_barchart.csv}; \label{plot:rpl_DS_multi}

    \addplot+[ybar, bar width=4pt,fill, color=black!50, error bars/error bar style={black}, area legend]  
            table [x expr= \thisrow{Holder}+0.25, y = RPL Dose, y error = RPL Error, col sep=comma] {plots/raw_data/rpl_barchart.csv}; \label{plot:rpl_bar_rpl}

    \addplot+[
        only marks,
        mark=triangle*,
        mark options={solid, scale=1.5}, 
        nodes near coords,
    ] coordinates {(0,0)};
    \label{DS_mark};

\end{axis}

\begin{axis}[
    name=box, 
    at={($(rpl.south east) + (\Hseparation,0)$)},
    boxplot/draw direction=y,
    anchor=south west,
    height=0.5\linewidth, 
    width=0.3\linewidth,
    xlabel=Dose-rate regime,
    ylabel=$\Delta D_{RCF-RPL}/D_{RPL}$ / \%,
    ymajorgrids=true,
    xmin=0.5, xmax=4.5,
    ymin=-27, ymax=27,
    xtick={1.5,3.5},
    xticklabels={UHDR, CONV},
    legend style={font=\footnotesize, 
        legend pos =north west},
    xticklabel style={
        font=\footnotesize, 
    },
    yticklabel style={
        font=\footnotesize, 
    },
    ]    
    \fill[gray, opacity=0.1] (axis cs:0.5,-27) rectangle (axis cs:2.5, 27); 

    \addplot+ [only marks, mark=*, violet!50]
    table [x expr=2*\coordindex/\coordindex, y = Film Multi UHDR, col sep=comma] {plots/raw_data/rpl_deviation.csv}; 

    \addplot+ [only marks, mark=-, mark size=6, line width=2,violet]
    table [x expr=2*\coordindex/\coordindex, y = Film Multi UHDR avg, col sep=comma] {plots/raw_data/rpl_deviation.csv}; 

    \addplot+ [only marks, mark=*,Chartreuse4!50]
    table [x expr=1*\coordindex/\coordindex,y = Film Green UHDR, col sep=comma] {plots/raw_data/rpl_deviation.csv}; 
    
    \addplot+ [only marks, mark=-, mark size=6, line width=2,Chartreuse4]
    table [x expr=1*\coordindex/\coordindex,y = Film Green UHDR avg, col sep=comma] {plots/raw_data/rpl_deviation.csv}; 

    \addplot+ [only marks, mark=*, violet!50]
    table [x expr=4*\coordindex/\coordindex, y = Film Multi CONV, col sep=comma] {plots/raw_data/rpl_deviation.csv}; 

    \addplot+ [only marks, mark=-, mark size=6, line width=2,violet]
    table [x expr=4*\coordindex/\coordindex, y = Film Multi CONV avg, col sep=comma] {plots/raw_data/rpl_deviation.csv}; 

    \addplot+ [only marks, mark=*,Chartreuse4!50]
    table [x expr=3*\coordindex/\coordindex,y = Film Green CONV, col sep=comma] {plots/raw_data/rpl_deviation.csv}; 
    
    \addplot+ [only marks, mark=-, mark size=6, line width=2,Chartreuse4]
    table [x expr=3*\coordindex/\coordindex,y = Film Green CONV avg, col sep=comma] {plots/raw_data/rpl_deviation.csv}; 

\end{axis}

\matrix [
            draw,
            fill=white,
            matrix of nodes,
            anchor=north west,
            at={($(rpl.south west) + (0*\Hseparation,-1.5)$)},
        ] {
             \ref{plot:rpl_bar_green} $\mathrm{RCF_{green}}$ & \ref{plot:rpl_bar_multi} $\mathrm{RCF_{multi}}$ & \ref{plot:rpl_bar_rpl} RPLD &
            \ref{plot:rpl_US_green} \ref{plot:rpl_US_multi} Upstream RCF & 
            \ref{plot:rpl_DS_green} \ref{plot:rpl_DS_multi} Downstream RCF\\
        };

\end{tikzpicture}
    \caption{\textbf{Left:} Bar chart of the individual dose measurements of the RPLDs and various RCF types analysed using both the single green channel and the multi-channel method. The chart display the average between upstream and downstream RCFs, as well as the upstream and downstream RCF measurements. The irradiations were performed at UHDR (holders 1-7) and CONV dose-rates (holders 8-16). \textbf{Right:} The deviations between RCF and RPLD measurements for each dose-rate regime. }
    \label{fig:rpl_results}
\end{figure*}
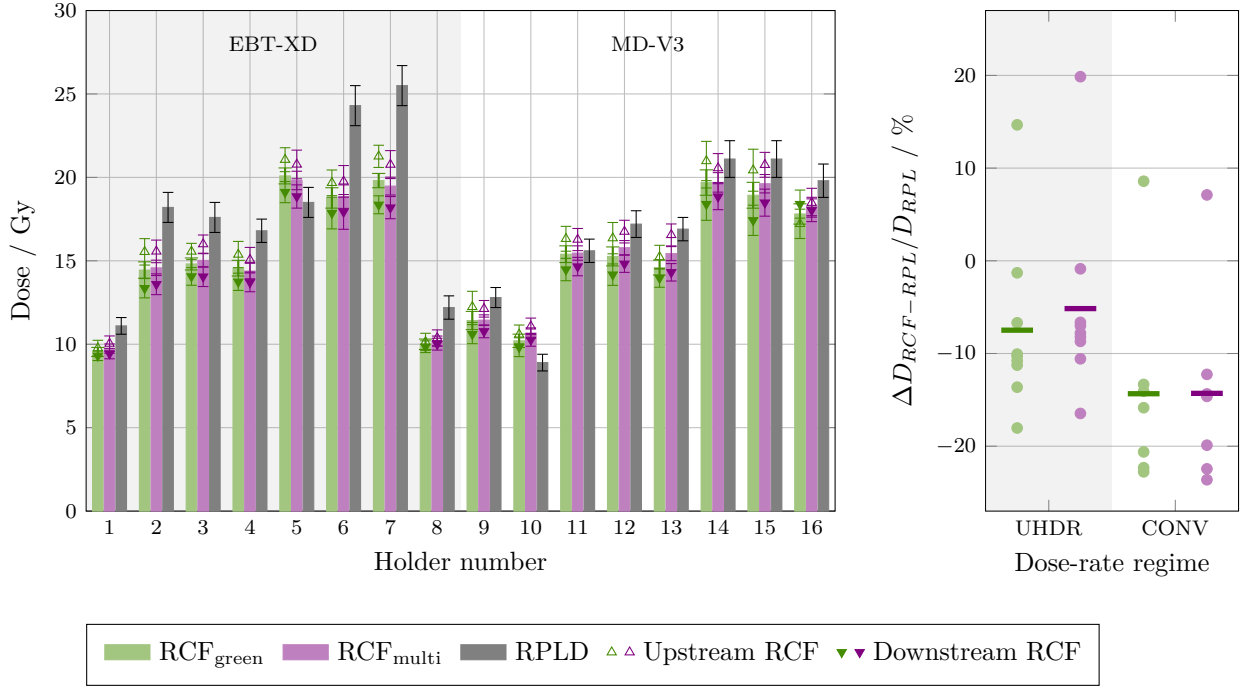

The RPLD doses were measured by the CERN radiation working group (RADWG). Both the single green channel and the multi-channel processing methods were used to evaluate the dose response of the RCFs. The mean doses and standard deviations were obtained from the ROI corresponding to the positions of the RPLDs and are shown in
Fig.~\ref{fig:rpl_results} along with the RPLD measurements. The left-hand plot shows that the dose values obtained from both the single green channel and multi-channel processing of the RCFs are generally lower and outside the standard deviations of the RPLD measurements. 

Overall, the mean standard deviation is lower for the RCF measurements (\SI{2.9}{\percent} and \SI{3.3}{\percent} for the multi-channel and green-channel, respectively) compared to the RPLDs (\SI{4.9}{\percent}). The left-hand plot also shows relative difference between RPLD and RCF is more variable than was the case for the ADs vs RCFs. The right-hand plot shows the single green channel and multi-channel RCF processing methods yield mean absolute relative deviations of more than \SI{5}{\percent} at UHDR and more than \SI{10}{\percent} at CONV with spreads of up to \SI{30}{\percent} relative to the RPLD measurements.

The significant and variable discrepancies observed between the RCFs and RPLDs can be attributed to multiple factors. Although the wide spread in relative measurements makes it challenging to draw definitive conclusions, there are several known issues that should be controlled for future experiments. First of all, the RPLDs were firmly inserted into a hole in a resin-printed sample holder, which raises the possibility that some residue may have rubbed off onto the RPLDs. Even though an ultrasonic bath cleaning procedure was performed, which resulted in a somewhat improved agreement with RCF measurements, it is possible that there may still be residues affecting the measurement. 

Secondly, although RPLDs are known to exhibit minimal dose rate effect, recent studies have indicated that exposure to dose rates differing from the calibration conditions can affect the formation of both RPL and colour centres within the glass~\cite{Aguiar2024DoseRange}.

Lastly, and most importantly, the calculation of relative dose between the RCFs and RPLDs is prone to misalignment errors. A slight offset between the region of the Gaussian beam intercepted by the RPLD and that analysed on the RCF would contribute to the relative dose uncertainty. Given that the cross-sectional area of the RPLDs is smaller than that of the ADs, alignment inaccuracies can have a more pronounced impact on the total absorbed dose recorded by the RPLDs.

\subsection{Comparison with Dosimetry Phantoms}

\begin{figure*}[htb!]
     \centering
     \input{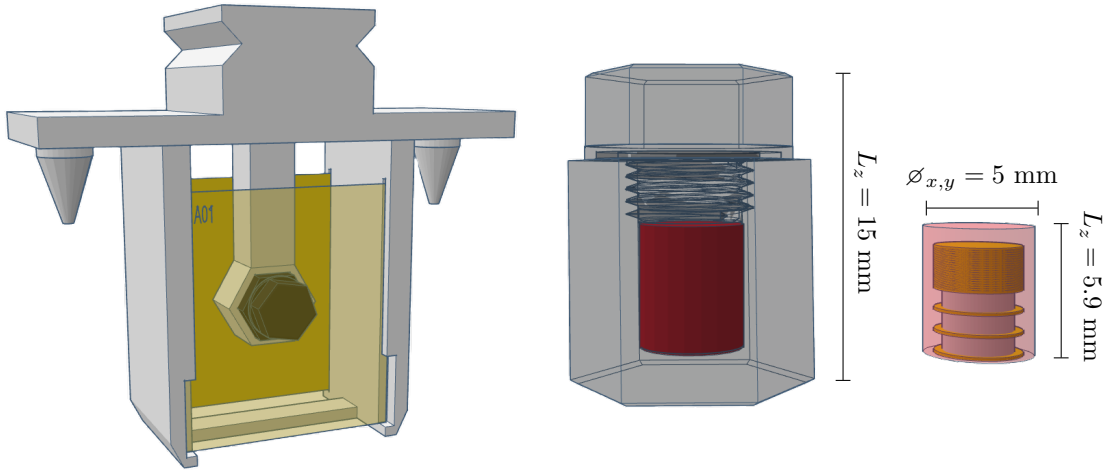}
    \caption{\textbf{Left:} The robot holder with the DP positioned between two RCFs. \textbf{Right:} The structure of the 3D printed DP.}
    \label{fig:phantom_setup}
\end{figure*}

Measurements comparing the dose response of the RCFs used in CLEAR and dosimetry phantoms (DPs) were performed, with the additional goal of assessing their relative agreement between CONV and UHDR dose-rates. The DPs were provided by the Institiut de Radiophysique (IRA) at CHUV in Switzerland. They are 3D-printed from ONYX$^{\circledR}$, a composite material of micro carbon fibre filled nylon with a density of \SI{1.2}{\gram\per\centi\metre\cubed}~\cite{onyx} and each contain three LiF-100 thermoluminescent dosimeters (TLD) interleaved with $\sim$~10 small ($\diameter=\SI{3}{\milli\metre}$) pieces of laser cut EBT3 RCFs. The remaining DP volume is filled with rubber to prevent large air gaps within the. The DPs were irradiated at CLEAR between two RCFs using the setup illustrated in Fig.~\ref{fig:phantom_setup}. This configuration is based on the multi-centre cross validation of the dosimetric comparison scheme for UHDR irradiation performed at CHUV and Stanford University, which used a cuboid DP filled with TLDs, ADs and RCFs to compare the measurements of \SI{8}{\mega\electronvolt} UHDR electron beams~\cite{Jorge2022DesignStudies}.

The setup is similar to that used during typical biological irradiations at CLEAR, which has an Eppendorf tube positioned between two RCFs. However the the DPs are relatively large compared to these tubes, and so to accommodate the length of DPs, the separation of the two RCFs was \SI{20}{\milli\metre} (vs \SI{12}{\milli\metre} for the tubes). We targeted \SI{10}{Gy} for all holders, which is a typical dose for biological irradiations at CLEAR. The agreement of different RCF types were simultaneously tested by interchanging between EBT3, EBT-XD and MD-V3 RCFs in the different sample holders. 

\begin{figure*}[htb!]
    \centering
    \pgfplotsset{compat=1.9}
\tikzset{
  mynode/.style={
  , align=center
  , execute at begin node=\setlength{\baselineskip}{1em}
  }
}

\begin{tikzpicture}
\def\Hseparation{2cm}
\def\Vseparation{1.2cm}

\begin{axis}[
    name=phantom, 
    anchor=south west,
    height=0.5\linewidth, 
    width=0.7\linewidth,
    ylabel={Dose~/~Gy},
    xlabel={Holder number},
    grid=both,
    xmin=0.5, xmax=10.5,
    xtick={1,2,...,10},
    xtick pos=bottom,
    ymin=0, ymax=15,
    error bars/y dir=both,
    error bars/y explicit,
    yticklabel style={
        font=\footnotesize, 
    },
    xticklabel style={
        font=\footnotesize, 
    },
    legend style={font=\footnotesize, 
        legend pos =south east},
    legend cell align={left},
    ]
    
    \fill[gray, opacity=0.1] (axis cs:0,0) rectangle (axis cs:6.5, 15); \node[align=center,anchor=center]at (axis cs:3.25,14){\footnotesize EBT3};

    \node[align=center,anchor=center]at (axis cs:7.5,14){\footnotesize EBT-XD};

    \fill[gray, opacity=0.1] (axis cs:8.5,0) rectangle (axis cs:10.5, 15); \node[align=center,anchor=center]at (axis cs:9.5,14){\footnotesize MD-V3};
    
    \addplot+[ybar, bar width=4pt,fill, color=Chartreuse4!50, error bars/error bar style={thick, Chartreuse4, mark size=2.5pt}, area legend]  
            table [x expr= \thisrow{Holder}-0.25, y = Film Green Avg, y error = Film Green Avg Err, col sep=comma] {plots/raw_data/phantom_barchart.csv}; \label{plot:phantom_bar_green}
    \addplot+[only marks, color=Chartreuse4, mark=triangle, forget plot, error bars/.cd, y dir=both, y explicit]
    table [x expr= \thisrow{Holder}-0.25, y = Film Green US, y error = Film Green US Err, col sep=comma] {plots/raw_data/phantom_barchart.csv}; \label{plot:phantom_US_green}
    \addplot+[only marks, color=Chartreuse4, mark=triangle*, mark options={rotate=180}, forget plot, error bars/.cd, y dir=both, y explicit]
    table [x expr= \thisrow{Holder}-0.25, y = Film Green DS, y error = Film Green DS Err, col sep=comma] {plots/raw_data/phantom_barchart.csv}; \label{plot:phantom_DS_green}
    \addplot+[ybar, bar width=4pt,fill, color=violet!50, error bars/error bar style={violet}, area legend]  
            table [x expr= \thisrow{Holder}-0, y = Film Multi Avg, y error = Film Multi Avg Err, col sep=comma] {plots/raw_data/phantom_barchart.csv}; \label{plot:phantom_bar_multi}
     \addplot+[xshift=-0pt, only marks, color=violet, mark=triangle, forget plot, error bars/.cd, y dir=both, y explicit]
    table [x expr= \thisrow{Holder}-0, y = Film Multi US, y error = Film Multi US Err, col sep=comma] {plots/raw_data/phantom_barchart.csv}; \label{plot:phantom_US_multi}
    \addplot+[xshift=-0pt,only marks, color=violet, mark=triangle*, mark options={rotate=180}, forget plot, error bars/.cd, y dir=both, y explicit]
    table [x expr= \thisrow{Holder}-0, y = Film Multi DS, y error = Film Multi DS Err, col sep=comma] {plots/raw_data/phantom_barchart.csv}; \label{plot:phantom_DS_multi}

    \addplot+[ybar, bar width=4pt,fill, color=black!50, error bars/error bar style={black}, area legend]  
            table [x expr= \thisrow{Holder}+0.25, y = Phantom Dose, y error = Phantom Error, col sep=comma] {plots/raw_data/phantom_barchart.csv}; \label{plot:phantom_bar_phantom}

\end{axis}

\begin{axis}[
    name=box,
    at={($(phantom.south east) + (\Hseparation,0)$)},
    boxplot/draw direction=y,
    anchor=south west,
    height=0.5\linewidth, 
    width=0.3\linewidth,
    xlabel=Dose-rate regime,
    ylabel=$\Delta D_{RCF-DP}/D_{DP}$ / \%,
    ymajorgrids=true,
    xmin=0.5, xmax=4.5,
    ymin=-17, ymax=27,
    xtick={1.5,3.5},
    xticklabels={UHDR, CONV},
    xticklabel style={
        font=\footnotesize, 
    },
    yticklabel style={
        font=\footnotesize,}, 
    legend style={font=\footnotesize, 
        legend pos =north west},
    ]    
    \fill[gray, opacity=0.1] (axis cs:0.5,-17) rectangle (axis cs:2.5, 27); 

    \addplot+ [only marks, mark=*,Chartreuse4!50]
    table [x expr=1*\coordindex/\coordindex,y = Film Green UHDR, col sep=comma] {plots/raw_data/phantom_deviation.csv}; 
    
    \addplot+ [only marks, mark=-, mark size=6, line width=2,Chartreuse4]
    table [x expr=1*\coordindex/\coordindex,y = Film Green UHDR avg, col sep=comma] {plots/raw_data/phantom_deviation.csv}; 
    
    \addplot+ [only marks, mark=*, violet!50]
    table [x expr=2*\coordindex/\coordindex, y = Film Multi UHDR, col sep=comma] {plots/raw_data/phantom_deviation.csv}; 

    \addplot+ [only marks, mark=-, mark size=6, line width=2,violet]
    table [x expr=2*\coordindex/\coordindex, y = Film Multi UHDR avg, col sep=comma] {plots/raw_data/phantom_deviation.csv}; 

    \addplot+ [only marks, mark=*,Chartreuse4!50]
    table [x expr=3*\coordindex/\coordindex,y = Film Green CONV, col sep=comma] {plots/raw_data/phantom_deviation.csv}; 
    
    \addplot+ [only marks, mark=-, mark size=6, line width=2,Chartreuse4]
    table [x expr=3*\coordindex/\coordindex,y = Film Green CONV avg, col sep=comma] {plots/raw_data/phantom_deviation.csv}; 

    \addplot+ [only marks, mark=*, violet!50]
    table [x expr=4*\coordindex/\coordindex, y = Film Multi CONV, col sep=comma] {plots/raw_data/phantom_deviation.csv}; 

    \addplot+ [only marks, mark=-, mark size=6, line width=2,violet]
    table [x expr=4*\coordindex/\coordindex, y = Film Multi CONV avg, col sep=comma] {plots/raw_data/phantom_deviation.csv}; 
\end{axis}

\matrix [
            draw,
            fill=white,
            matrix of nodes,
            anchor=north west,
            at={($(phantom.south west) + (0*\Hseparation,-1.5)$)},
        ] {

                \ref{plot:phantom_bar_green} $\mathrm{RCF_{green}}$ & \ref{plot:phantom_bar_multi} $\mathrm{RCF_{multi}}$ & \ref{plot:phantom_bar_phantom} DP &
                \ref{plot:phantom_US_green} \ref{plot:phantom_US_multi} Upstream RCF & 
                \ref{plot:phantom_DS_green} \ref{plot:phantom_DS_multi} Downstream RCF\\
        };

\end{tikzpicture}
    \caption{\textbf{Left:} Bar chart of dose measurements from each irradiation holder of the DPs and various RCF types analysed using both the single green channel and the multi-channel method. The chart display the average between upstream and downstream RCFs, as well as the upstream and downstream RCF measurements. The irradiations were performed at UHDR (holders 1\textendash 3, 7 and 9) and CONV (holders 4\textendash 6, 8 and 10) dose rates. \textbf{Right:} The deviations between RCF and DP measurements for each dose-rate regime.}
    \label{fig:phantomdoses}
\end{figure*}
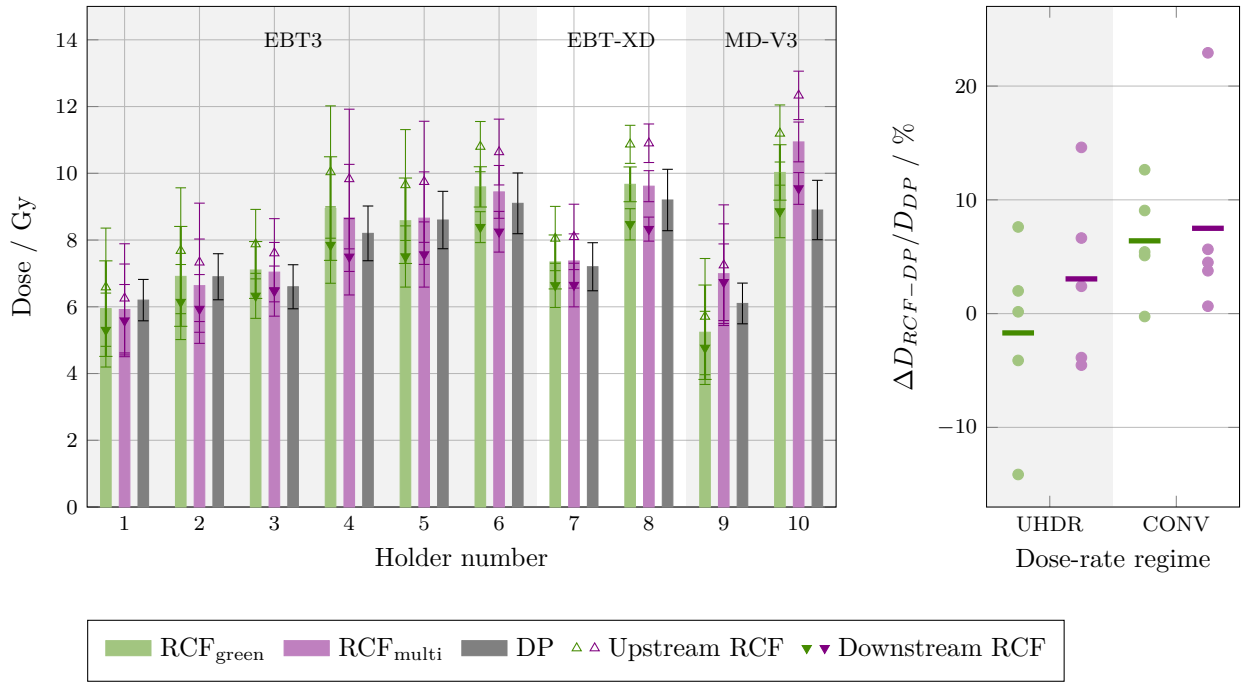

The DP doses were evaluated at IRA, and are represented as the mean and standard deviation across measurements from all dosimeters inside the DP. Both the single green channel and the multi-channel processing methods were used to evaluate the dose response of the upstream and downstream RCFs. The mean doses and standard deviations were obtained from the ROI corresponding to the positions of the cross-section of the sensitive volume of the DPs. These results are shown in
Fig.~\ref{fig:phantomdoses}. 

The left-hand plot shows that, for most of the holders, the dose obtained from the DP is within the uncertainty range of the mean RCF response. The magnitude of the individual standard deviations can be attributed to the small (Gaussian) beam size relative to the cross-section of the DPs, resulting in a large transverse variation in dose.
Moreover, the significant beam scattering over the distance between the two RCFs is the reason for the notable differences in the doses measured by the upstream and downstream RCFs\textemdash and potentially one of the reasons for discrepancies in the doses determined by the RCFs and the DPs. It is also clear that the relative difference the upstream and downstream RCFs is variable. This could indicate that the beam divergence is inconsistent between irradiations, making it difficult to evaluate the accuracy of the measurements. 

The right-hand plot shows the relative deviation between RCFs and DPs for UHDR and CONV. There seems to be a higher spread for measurements at UHDR versus CONV. This could potentially be attributed to pulse-to-pulse position jitter which is more prominent ($\simeq \SI{1}{\milli\metre}$) for measurements at UHDR compared to CONV (where several pulses are accumulated). This error in alignment between the sample and the centre of the beam increases the gradients across the DPs which can increase the uncertainty significantly.

In addition to the reasons discussed above, there are a couple of known uncertainties which may in part explain the discrepancies in the results. One is that due to an issue with the accelerator, half of th DPs were irradiated \SI{13}{\day} after the other half. At IRA, the protocol is to irradiate reference TLDs in a $^{60}\mathrm{Co}$ beam to a known reference dose the same day as the irradiation of the TLDs being tested. This allows for correction of the daily sensitivity of the entire dosimetric system and thus avoids the need for additional fading corrections~\cite{jaccard_RCF_DR}.
Because the reference irradiation was performed on the first day, this could imply that the uncertainty of the last 5 phantom measurements are higher.

\section{Conclusions}
This paper presented an evaluation of the various factors that impact RCF dosimetry.  This was used to motivate the RCF dosimetry protocol that has been implemented in the CLEAR facility to ensure that desired precision is optimised with feasibility in term of time and resources. This protocol is particularly useful for VHEE/UHDR research facilities that with a high throughput of RCFs.

The analysis presented shows that establishing and strictly adhering to a protocol is imperative for RCF dosimetry. To ensure accuracy, there must be complete consistency in the handling of both calibration RCFs and application RCFs. For example, they must be from the same production lot, be stored in the same conditions, and analysed by applying the same scanning procedure. The latter includes physical considerations such as RCF orientation and position, using the same digitiser that has been sufficiently warmed up,  as well as logistical considerations such as ensuring that the scanner plate and RCF are both free from dust and stains and scanning at a consistent post-exposure time. Failure to adhere to such protocol introduces errors and, for example, we showed that an inconsistent timing between  scans post-exposure can  result in a dose uncertainty of \SI{5}{\percent}. We particularly emphasise this point as it is often overlooked at a facility such as CLEAR which does not have a reference beam for re-calibration on site. Moreover, for facilities with small and non-uniform beams in particular, it is important be aware that the presence of e.g. water residues can significantly alter the dose measurement. All these effects have been described and, when possible, quantified in this paper.

The paper presented the estimates of expected accuracies of using the CLEAR RFDS for dosimetry for irradiations, by comparison of RCF measurements to ADs, RPLDs, and DPs. We showed a relative agreement within \SI{5}{\percent} compared with ADs using a Gaussian beam whereas the agreement was lower between PRDLs and PDs. We explained the latter inconsistencies were  likely due to experimental uncertainties such as alignment inaccuracy, which is a challenge for small beams. By leveraging the experience gained from these results alongside recent advancements in irradiation procedures at CLEAR, future optimised experiments are foreseen and expected to yield a better agreement. In particular, keeping the distance between RCF and sample as small as possible helps reducing uncertainties. Moreover, for small Gaussian beams, a precise alignment and RCF ROI selection is crucial. Lastly, and particularly for irradiations at UHDR, a better control on pulse-to-pulse beam position would likely improve the results.

\section*{Acknowledgements}
This work has to a great extent been based on the comprehensive compilation of best-practices for RCF dosimetry by Niroomand-Rad et al. in the Report of AAPM Task Group 235~\cite{rcf_composition}, the numerous papers of S. Devic \textit{et al.} which thoroughly describe various aspects of RCF dosimetry ~\cite{rcf_history, devic_digitizers, devic_precise, devic_range, Devic2006SensitivityDosimetry}, and the literature on multichannel dosimetry by Micke et al.~\cite{micke_multi} and Mayer et al.~\cite{mayer_multi}. Moreover, we have to extend our gratitude towards CHUV, HUG, PTB, and the CERN radiation working group (RADWG), for their support and useful input in calibrations and dosimetry cross-correlation experiments, which has been instrumental in this work.  

\bibliographystyle{mynum.bst}
\bibliography{mendeley}

\end{document}